\documentclass[useAMS,usenatbib]{mn2e}
\usepackage{graphicx}
\newcommand{\apj}{ApJ}

\newcommand{\aap}{A\&A}

\newcommand{\aj}{AJ}
\newcommand{\mnras}{MNRAS}
\newcommand{\nat}{Nature}
\newcommand{\pasp}{PASP}
\newcommand{\apjs}{ApJS}
\newcommand{\MC}{\multicolumn}

\DeclareRobustCommand{\ion}[2]{%
\relax\ifmmode
\ifx\testbx\f
{\mathrm{#1\,\textsc{#2}}}\else
{\mathrm{#1\,\mathsc{#2}}}\fi
\else\textup{#1\,{\mdseries\textsc{#2}}}%
\fi}
\newcommand{\sunn}{$_{\odot}$}
\newcounter{qub}
\title[The faint outer regions of the Pegasus Dwarf Irregular Galaxy]
{The faint outer regions of the Pegasus Dwarf Irregular galaxy:
a much larger and undisturbed galaxy
}

\author[A. Y. Kniazev et al.]{%
Alexei Y. Kniazev,$^{1,2}$\thanks{E-mail: akniazev@saao.ac.za (AYK)}
Noah Brosch,$^{3}$
G. Lyle Hoffman,$^{4}$
Eva K. Grebel,$^{5}$
\newauthor
Daniel B. Zucker,$^{6,7,8}$
Simon A. Pustilnik$^{9,10}$\\
\rule{-4pt}{20pt}
$^{1}$South African Astronomical Observatory, PO Box 9, 7935, Cape Town, South Africa\\
$^{2}$Southern African Large Telescope Foundation, PO Box 9, 7935, Cape Town, South Africa\\
$^{3}$The Wise Observatory and the Raymond and  Beverly Sackler School of Physics
and Astronomy, the Faculty of Exact Sciences, \\ Tel Aviv University, Tel Aviv 69978, Israel\\
$^{4}$Department of Physics, Lafayette College, Easton, PA 18042, USA \\
$^{5}$Astronomisches Rechen-Institut, Zentrum f\"ur Astronomie der Universit\"at Heidelberg,
M\"onchhofstr.\ 12--14, D-69120 Heidelberg, Germany\\
$^{6}$Institute of Astronomy, University of Cambridge, Madingley Road, Cambridge CB3 0HA,
United Kingdom\\
$^{7}$Department of Physics, Macquarie University, North Ryde, NSW 2109, Australia\\
$^{8}$Anglo-Australian Observatory, PO Box 296, Epping, NSW 1710, Australia\\
$^{9}$Special Astrophysical Observatory, Nizhnij Arkhyz, Karachai-Circassia, 369167, Russia\\
$^{10}$Isaac Newton Institute of Chile, SAO Branch, Nizhnij Arkhyz, Russia
}

\begin{document}

\date{Accepted 2009 August ??. Received 2009 July 31; in original form 20?? October ??}

\pagerange{\pageref{firstpage}--\pageref{lastpage}} \pubyear{2009}

\maketitle

\label{firstpage}

\begin{abstract}
We investigate the spatial extent and structure of the Pegasus
dwarf irregular galaxy using deep, wide-field, multicolour CCD
photometry from the Sloan Digital Sky Survey (SDSS) and new deep HI observations.
We study an area of $\sim$0.6 square degrees centred on the Pegasus dwarf that
was imaged by SDSS. Using effective filtering in
colour-magnitude space we reduce the contamination by
foreground Galactic field stars and increase significantly the
contrast in the outer regions of the Pegasus dwarf.

Our extended surface photometry, reaches down to a surface brightness
magnitude $\mu_r\simeq32$ mag arcsec$^{-2}$.
It reveals a stellar body
with a diameter of $\sim$8 kpc
that follows a S\'{e}rsic surface brightness distribution law,
which is composed of a significantly older stellar population than that
observed in the $\sim$2 kpc main body.
The galaxy is at least five times more extended than
listed in NED. The faint extensions of the galaxy are not equally
distributed around its circumference; the north-west end is more
jagged than the south-east end. We also identified a number of
stellar concentrations, possibly stellar associations, arranged in a ring
around the main luminous body.

New HI observations were collected at the Arecibo Observatory
as part of the ALFALFA survey.
They reveal an HI distribution somewhat elongated in RA and about 0\fdg3 wide,
with the region of highest column density coincident with the luminous galaxy.
The HI rotation curve  shows a solid-body rotation behaviour,
with opposite ends differing by 15 km s$^{-1}$.
There is a stream to lower
velocities about 5 arcmin from the centre of the galaxy.

We were able to measure $ugriz$ colours in a number of apertures using the SDSS
data and compared these with predictions of evolutionary synthesis models.
The results indicate that the outermost regions of PegDIG are
5--10 Gyr old, while the inner kpc contains stars $\sim$1 Gyr
old and younger. The colours correspond to K-stars;
earlier subclasses are located in the innermost parts of the galaxy.
PegDIG appears to be a relatively low-mass object, with a total dynamical
mass of 3$\times10^8$ M$_{\odot}$ of which only 30\% in stars and 2\%
is in neutral gas.

The extended stellar distribution,
the appearance of faint light extensions,
and the lack of low column density HI tails rule out a possible tidal
origin or a ram pressure stripping scenario. We propose that PegDIG
is a fairly recent acquisition by the Local Group,
since it does not appear to be disturbed by interactions with other galaxies.
\end{abstract}


\section{Introduction}

The Local Group (LG), our immediate neighbourhood, is the group of galaxies
in which our Milky Way galaxy formed and is evolving. The Milky Way and M31
are the two dominant spiral galaxies in the Local Group, and each is surrounded
by an entourage of lower-mass companions.  This
kind of a binary structure is found in nearby galaxy groups as well
\citep[e.g,][]{Kara02a,Kara02b,Kara03}. Apart from the late-type spiral M33,
the more than fifty galaxies forming the LG are either dwarf elliptical (dE),
dwarf spheroidal (dSph) or dwarf irregular (dIrr) galaxies.
\citet{Greb01} characterised galaxies as dwarf galaxies if their total B-band
or V-band magnitude is fainter than $-18$ mag. While the primary distinguishing
feature between dIrrs and dSphs is whether they have or do not have HI,
there are other distinguishing features. dIrr galaxies typically have central
surface brightnesses of $\mu_{V, 0}\leq$ 23 mag arcsec$^{-2}$ and
total H\,{\sc i} masses $M_{HI} \leq 10^9$ M$_{\odot}$. As their name suggests,
their optical appearance is irregular, which tends to be caused by scattered
H\,{\sc ii} regions for the more massive dIrrs.  They are mainly found at
larger distances from massive galaxies. dEs are spherical or elliptical
in appearance, typically with M$_V\geq$--17 mag, and have $\mu_{V, 0}\leq$ 21
mag arcsec$^{-2}$ and $M_{HI} \leq 10^8$ M$_{\odot}$. Along with the dSphs, they are
usually found in the vicinity of massive galaxies.  dSphs tend to
have M$_V\geq$--14 mag, $\mu_{V, 0}\geq$ 22 mag arcsec$^{-2}$,
and $M_{HI} \leq 10^5$ M$_{\odot}$. See \citet{Greb01} for further details.

Dwarf galaxies are, in principle, simpler systems than large galaxies.
However, those objects studied in detail showed that even dwarf systems
are complex in terms of their extended star formation histories and
chemical evolution \citep[e.g,][]{Greb97,Mat98}.  Moreover, in the LG
there is evidence for past interactions among some of its galaxies,
including the accretion of dwarf galaxies by the two large spirals
\citep[e.g,][]{IGI94,Ibata01,Yanny03,Zucker04a,Belokurov07a,Bell08}.

The LG member galaxies reside currently in a volume with a
radius of $\sim$ 1 -- 1.2 Mpc and their integrated mass is
estimated to be between ($1.3 \pm 0.3$)$\times 10^{12}$ to
(2.3$\pm$0.6)$\times 10^{12}$ M$_{\odot}$ \citep{CvdB99,Kara02c}.
The LG is the best-studied galaxy group thus far and includes the faintest
dwarf galaxies ever detected \citep[e.g,][]{Zucker06,Belokurov06,Belokurov07b}.
The dwarf, low-luminosity bodies in the LG  all contain old
populations \citep{GreGal04} and may represent relics from the re-ionisation
epoch \citep{GK06}.
Finally, there are persistent claims that dwarf galaxies are
dark-matter (DM) dominated. Studying their properties in the LG,
the place where the most detailed work can be done, is a step toward
understanding the nature of DM \citep[e.g.,][]{Gilmore07}.

It is important to determine how large a galaxy is in any observational band.
The apparent size alters the baryonic mass estimate and shows how
extended the stellar component is in comparison to the gas.
This has also implications on the possible
evolution of the object, because the larger a galaxy's cross-section,
the higher are its chances of having interacted with other
galaxies since its formation. The shape of the outer regions of a
galaxy is a particularly sensitive indicator of past interactions;
these tend to strip away stars and ISM creating tidal tails and
spreading debris in the vicinity of a galaxy. However, it
is very difficult to determine the morphology of the outer
extremely faint surface brightness regions for galaxies in the LG.
Their closeness implies a large angular extent.
If the coverage of faint surface brightness levels over large areas
is required, the necessary observing time for direct, wide-field
imaging becomes prohibitive.
In addition, it is important to note that distorted galaxy shapes at
low light levels can be an artefact of Poisson scatter \citep{M08}.

The availability of imaging data from the Sloan Digital
Sky Survey \citep[SDSS,][]{York2000,Stoughton02}
offers a new and different way to approach this task. This data
set is very uniform, covers a large fraction of the sky in five
spectral bands, reaches objects as faint as 24 mag, and is
publicly available. The SDSS imaging data have already used to
explore the structural parameters of other
LG objects (e.g., Draco: \citet{Odenk01b},
And IX: \citet{Zucker04b}; And X: \citet{Zucker07};
Ursa Major: \citet{Willman05}; Leo\,II: \citet{Coleman07},
and many others).

In this paper we present a new interpretation of observational material
of the Pegasus dwarf irregular galaxy (DDO 216=UGC 12613, hereafter referred
to as PegDIG) based on very deep and wide-field surface photometry using imaging
data from the SDSS. PegDIG is one of the fainter LG members
and contains very few stars younger than 100 Myr \citep[see also ][]{Gall98}.
This object is not to be confused with the dwarf spheroidal galaxy
Pegasus II = Peg dSph = Andromeda VI \citep{KK99,GreGu99,Armand99}.

We trace PegDIG across its entire angular extent in the
optical to the lowest accessible surface
brightness levels using SDSS data, since all previous studies
concentrated either on stars near the centre of the galaxy
\citep[e.g.,][]{Gall98} or presented surface brightness profiles
covering only small fields \citep[e.g.,][]{Zee00}.
We combine these data with new H\,{\sc i} maps derived from
Arecibo Observatory\footnote{%
The Arecibo Observatory is part of the National Astronomy and Ionosphere Center,
which is operated by Cornell University under a cooperative agreement with
the National Science Foundation.} data collected for the ALFALFA
survey \citep[e.g.,][]{Haynes08}.

\section{The Pegasus dwarf irregular galaxy}
\label{txt:summary_for_Peg}

As discussed by \citet{Gall98}, PegDIG exhibits many of the properties
typical of dIrr galaxies such as H\,{\sc i}, young stars, and H\,{\sc ii}
regions, and an irregular appearance due to recent star formation.  On
the other hand, it is at the faint end of the luminosity range of dIrrs,
its outer isophotes are fairly smooth, its gas content
is comparatively low, and its recent star formation activity is low as
well, prompting \citet{Gall98} to suggest that it may be a dIrr/dSph
transition-type galaxy.

\subsection{Optical data, morphology, star formation and distance estimates}

The basic parameters of PegDIG, compiled from the literature, are listed in
Table~\ref{tbl:General}. There has been significant disagreement among the various distance
estimators presented in the past, ranging from 1.75 Mpc derived from Cepheids to 760 kpc
derived from the tip of the red giant branch (TRGB) in colour-magnitude diagrams.
\citet{dV75} quoted a distance of only 170 kpc to PegDIG, considering the galaxy one
of our closest neighbours. The three most recent determinations all use ground-based
photometric data and employ the TRGB \citep{McC05,Ti06} or Cepheids \citep{MGC09}
as distance indicators.  These studies converge to a value of about 1 Mpc,
slightly larger than the distance of $760 \pm 100$ kpc derived by \citet{Gall98}
using data from the Hubble Space Telescope.
These distances render PegDIG a peripheral, potential member of the M31 subgroup
within the LG. We adopt here a nominal distance to PegDIG of 1 Mpc.

\begin{table*}
\caption{Basic Parameters of the Pegasus Dwarf Irregular Galaxy \label{tbl:General}}
\begin{tabular}{lllll} \hline
\MC{1}{c}{Parameter}    & \MC{3}{c}{Value}                            & \MC{1}{c}{Reference}        \\
\MC{1}{c}{(1)}          & \MC{3}{c}{(2)}                              & \MC{1}{c}{(3)}              \\ \hline
$\alpha$ (J2000.0)      & \MC{3}{c}{23$^h$28$^m$36.25$^s$}            &                             \\
$\delta$ (J2000.0)      & \MC{3}{c}{+14$^\circ$44$^\prime$34.5$^{\prime\prime}$} &                  \\
B$_T$    (mag)          & \MC{3}{c}{12.50$\pm$0.15}                   &   \citet{Zee00}             \\
$(B-V)$  (mag)          & \MC{3}{c}{0.64$\pm$0.04}                    &   \citet{Zee00}             \\
$(U-B)$  (mag)          & \MC{3}{c}{0.00$\pm$0.09}                    &   \citet{Zee00}             \\
$\alpha_B$ (arcsec)     & \MC{3}{c}{104.4}                            &   \citet{Zee00}             \\
\\
Distance estimates     &        &                & &
\\ \hline
	    &Method  &  Distance Modulus     &     kpc           & \MC{1}{c}{Reference}        \\
	    &\MC{1}{c}{(1)}&\MC{1}{c}{(2)}&\MC{1}{c}{(3)}&\MC{1}{c}{(4)}                \\ \hline
	    & PN     &                & 200--1600         &   \citet{JL81}              \\
	    & TRGB   & $\approx$26    &  1700             &   \citet{HM82}              \\
	    & TRGB   & $\approx$25.5  &  700--1200        &   \citet{CT83}              \\
	    &Cepheids&26.1$\pm$0.3    &  1700             &   \citet{Hoessel90}         \\
	    & TRGB   & 24.9$\pm$0.1   &  950$\pm$50       &   \citet{Ap94}              \\
	    & TRGB   & 25.13$\pm$0.11 &  1060$\pm$50      &   \citet{Lee95}             \\
	    &TRGB, RC& 24.40          &  760$\pm$100      &   \citet{Gall98}            \\
	    & TRGB   & 24.80$\pm$0.07 &  919$\pm$30       &   \citet{McC05}             \\
	    &TRGB    & 25.04          &  1020$\pm$20      &   \citet{Ti06}              \\
	    &Cepheids&                &  1070$\pm$50      &   \citet{MGC09}             \\ \hline
\MC{5}{p{15.5cm}}{%
Note: The abbreviations used here are as follows: $\alpha_B$ is the exponential scale
length of the disk fitted by \citet{Zee00} to the $B$ surface brightness profile;
PN refers to a distance estimate using planetary nebulae; TRGB is a distance using the
tip of the red giant branch in the colour-magnitude diagram; RC is the same, but using
the ``red clump''.}
\end{tabular}
\end{table*}


\begin{figure}
    \begin{center}
    \includegraphics[angle=0,width=8.5cm,clip=,]{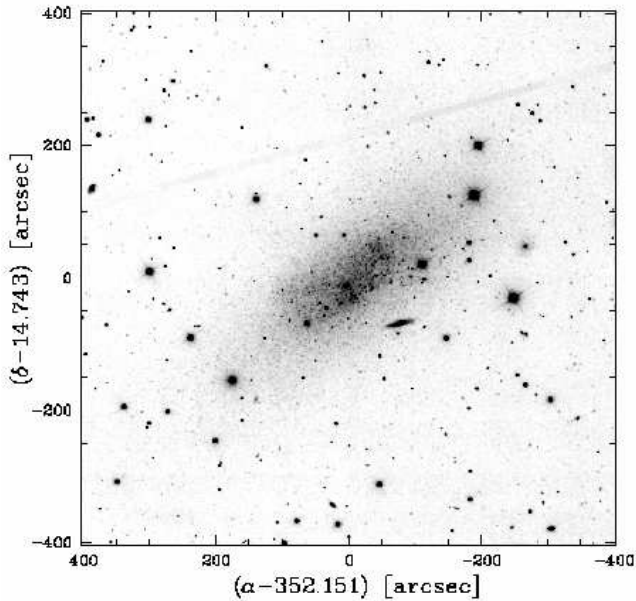}
    \caption{
     The central part ($\sim13.5\times13.5$ arcmin) of combined $g, r$, and $i$
     image of Pegasus dwarf galaxy derived from a weighted combination
     of individual single-colour images from the SDSS.
     North is up and East is to the left.
     The oblique faint line is a satellite track that was
     registered on the SDSS $r$ image.
     At the adopted distance of 1.0 Mpc, 1\arcsec\ = 4.85 pc.
    \label{fig:Direct_image}}
    \end{center}
\end{figure}

\citet{Gall98} studied the central region of PegDIG using
imaging in the B, V and I bands with the Wide Field and Planetary
Camera (WFPC2) aboard the Hubble Space Telescope (HST).  Their
field encompasses regions of recent SF in PegDIG.  These authors
obtained the deepest photometry of PegDIG so far, reaching objects
down to V$\simeq$25. They identified a main sequence as well as blue
loop stars younger than 0.5 Gyr.  These young populations are
clustered in two central
clumps. Moreover, there are older, more widely distributed stars.
\citet{Gall98} and \citet{D-P98} concluded that the model best
fitting their measurements has a relatively constant SF rate
over several Gyrs, with a period of enhanced SF about 2 Gyrs ago.
The colours of the main-sequence stars measured by \citet{Gall98}
indicate a relatively high extinction of
A$_V\simeq$0.47 or $E(B-V) = 0.15$, and the stellar models could match
the observations only for a distance to PegDIG of 760 kpc.
\citet{Gall98} combined various measurements including the H\,{\sc i}
column density and IRAS dust maps in order to constrain the reddening
and pointed out that these estimators do not yield consistent results.
The adopted extinction is one of the main reasons for their
shorter distance.

\citet{Krienke01} used the colours and redshifts of background
galaxies visible in the HST/WFPC2 data of \citet{Gall98}
to estimate the internal reddening in the central region of PegDIG.
After correcting for the Galactic foreground reddening using the values of
\citet{Schlegel98}, they found that PegDIG itself contributes very
little to the reddening and seems to have hardly any interstellar
dust.  \citet{Krienke01} derived an internal reddening of $E(B-V) \leq
0.03$ for PegDIG, consistent with \citet{Gall98}.  The total reddening
of \citet{Krienke01} is lower than that of \citet{Gall98} by $\geq 0.05$
in $E(B-V)$.

\citet{YvZL03} estimated a current star formation rate of
3$\times10^{-5}$ M$_{\odot}$ yr$^{-1}$ from the H$\alpha$ flux,
based on the broadband and H$\alpha$ imaging in \citet{Zee00}.
\citet{HE04} listed PegDIG as having the lowest star formation
rate among the 94 irregular galaxies of their sample
($4.4\times10^{-5}$ M$_{\odot}$ yr$^{-1}$) and the lowest star
formation rate per surface area ($8.3\times10^{-4}$ M$_{\odot}$
yr$^{-1}$ kpc$^{-2}$).

NED lists a major axis of 5 arcmin
($\sim$1.5 kpc at 1 Mpc) and a minor axis of 2.7 arcmin for PegDIG.
\citet{Ti06} found a 5 kpc thick disk (diameter $\sim$16 arcmin at 1
Mpc) by tracing the red giant stars on images obtained with the
6-m BTA telescope and with HST. He wrote that the
low-luminosity blue stars in PegDIG are scattered throughout the
5\arcmin$\times$3\arcmin body of the galaxy. The HST image allowed the
decomposition of the blue star distribution perpendicular to the
disk; these stars ``virtually disappear at 200 pc'' while the
thick disk can be traced vertically to $\sim$1 kpc.

Recently, \citet{MVI07} compared the stellar structure
of PegDIG derived from Johnson V and Gunn $i$ images from the
INT Wide Field Camera  with the H\,{\sc i} maps produced by \citet{YvZL03}.
They concluded that the H\,{\sc i} distribution is different from
that of the ``regular, elliptical'' stellar distribution, and
interpreted that as strong evidence for ram-pressure stripping of
the PegDIG by an intergalactic medium associated with the Local
Group.

\subsection{The interstellar matter:  HI distribution, mass, and dynamics}

\citet{FT75} were the first to detect H\,{\sc i} emission from
PegDIG and, based on its low radial velocity, suggested it is a
Local Group member. The 21-cm line emission they measured had a
width of 43$\pm$7 km sec$^{-1}$.
The neutral hydrogen in PegDIG was mapped by \citet*{LSY93} with
the VLA, and by \citet{HSF96} with the Arecibo radio
telescope. \citet{LSY93} showed that the H\,{\sc i} is concentrated in the
main optical body and that it shows some clumps. \citet{HSF96}
found that the H\,{\sc i} distribution was asymmetric, with the H\,{\sc i}
peak offset by $\sim 1'$ from the centre of the outer
iso-intensity contour. The rotation curve derived from the Arecibo
observations becomes flat at about $\sim$4.5 arcmin SE of the kinematic
centre, which is itself offset by $\sim$2.6 arcmin from the
optical centre. On the NW side the rotation curve falls off after
peaking at $\sim$2 arcmin from the kinematic centre. Note that
\citet{YvZL03} mention that the low velocity of PegDIG
very close to zero -- causes significant confusion with Galactic
H\,{\sc i}, thus the interpretation of single-dish wide-beam H\,{\sc i} measurements
of this galaxy may be problematic.

\citet{Kar01} observed PegDIG with the Effelsberg
radio telescope as part of a program to map the local galaxy
population. They give only an upper limit of 11 mJy from the H\,{\sc i}
line flux, which they attribute to a low H\,{\sc i} content and not to
confusion from local H\,{\sc i}. Note that the 21-cm half-power beam at
Effelsberg is 9.3 arcmin.

\begin{table*}
    \begin{center}
    \caption{Photometric parameters for the unresolved part of PegDIG in
    the SDSS
    \label{tbl:Intergral}}
    \begin{tabular}{clcrllll} \hline
   \MC{1}{c}{Filter}
   & \MC{4}{c}{Circular Apertures}
   & \MC{3}{c}{2D GALFIT}\\ \hline
   & \MC{1}{c}{b/a}
   & \MC{1}{c}{$\mu_0$}
   & \MC{1}{c}{$\alpha$}
   & \MC{1}{c}{$n$}
   & \MC{1}{c}{b/a}
   & \MC{1}{c}{$r_e$}
   & \MC{1}{c}{$n$}
   \\
   &
   & \MC{1}{c}{[mag/sq.arcsec]}
   & \MC{1}{c}{[arcsec]} &
   &
   & \MC{1}{c}{[arcsec]}
   & \\
    \MC{1}{c}{(1)}&\MC{1}{c}{(2)}&\MC{1}{c}{(3)}  &\MC{1}{c}{(4)}& \MC{1}{c}{(5)} &\MC{1}{c}{(6)}&\MC{1}{c}{(7)}  &\MC{1}{c}{(8)} \\ \hline
    $u$           & 0.40         & 24.24$\pm$0.10 & 89$\pm$9     & 1.05$\pm$0.05  &0.40$\pm$0.01&  172:    ~~(98:)&0.96$\pm$0.04  \\   
    $g$           & 0.40         & 23.10$\pm$0.05 & 98$\pm$5     & 1.03$\pm$0.02  &0.40$\pm$0.01&  172$\pm$1 (108)&1.04$\pm$0.02  \\   
    $r$           & 0.40         & 22.73$\pm$0.04 & 114$\pm$5    & 1.10$\pm$0.02  &0.41$\pm$0.01&  172$\pm$1 (126)&1.18$\pm$0.02  \\   
    $i$           & 0.40         & 22.46$\pm$0.05 & 118$\pm$5    & 1.11$\pm$0.03  &0.39$\pm$0.01&  177$\pm$1 (139)&1.25$\pm$0.02  \\   
    $z$           & 0.40         & 22.40$\pm$0.07 & 127$\pm$7    & 1.33$\pm$0.06  &0.40$\pm$0.01&  164$\pm$2 (125)&1.22$\pm$0.02  \\   
    \hline
    \end{tabular}
    \end{center}
\end{table*}

\citet{Zee00} reported results of a study of isolated dwarf
galaxies, where ``isolated'' implies a distance of 100--200 kpc
from the nearest neighbour, using UBV and H$\alpha$ imaging,
combined with VLA H\,{\sc i} mapping. PegDIG was included in her sample
and yielded an integrated flux of 29.90 Jy km s$^{-1}$ and an
H\,{\sc i} line
width of 40 km s$^{-1}$ for a recession velocity of $-183$ km
s$^{-1}$, based on the VLA observations reported in \citet{YvZL03}.
van Zee's (2000) photometry showed that PegDIG has the reddest
colours of all galaxies in her sample. Evolutionary
synthesis models indicate that the stellar population could be
the result of a single major star formation (SF) burst a few Gyr
ago that ran out of material for further SF.

Further H\,{\sc i} mapping of PegDIG was done at Westerbork by \citet{SI02},
as part of a project to study the hydrogen in a
sample of dwarf galaxies. PegDIG was considered by these authors as one of
their most isolated objects. \citet{SI02} found a 9.0 mJy upper
limit to the 1.4 GHz continuum flux from the galaxy in the 1.35 MHz
band, and a 21-cm line flux integral of 16.3$\pm$0.5 Jy km
s$^{-1}$, implying a total H\,{\sc i} content of 4$\times10^6$
M$_{\odot}$ at the assumed 1 Mpc distance. The map they
presented in their Fig.\ 4 shows the H\,{\sc i} concentrated at
$\alpha(2000)=23^h26^m04^s$ and $\delta(2000)=+14\degr27^m40^s$,
although the $13''$ synthesised beam was rather
elongated in the north-south direction because of the declination
of the object. The line profile shown in their Fig.\ 2 is
single-peaked and very narrow, with a FWHM of less than 20 km
s$^{-1}$ and a weak $\sim80$ km s$^{-1}$ tail to higher
recession velocities (the systemic H\,{\sc i}  velocity of the galaxy
being $-189$ km s$^{-1}$).

The most recent published H\,{\sc i} map of PegDIG was by \citet{YvZL03}
using the VLA and combining observations performed in the
C and D configurations. Their integrated 21-cm  profile has a
50\% full width of 24.6 km s$^{-1}$ and a flux integral of 29.9
Jy km s$^{-1}$. \citet{YvZL03} display in their Fig.\ 5 a contour
map of PegDIG with the lowest column density level at 10$^{19}$
atoms cm$^{-2}$ and a $\sim$10 arcmin extent that shows the gas
arranged in three main clumps. In addition, a region 1.5 arcmin
northwest of the centre shows a double H\,{\sc i} profile; this is
interpreted as an expanding bubble with a radius of one arcmin
($\simeq$200 pc at the 760 kpc assumed distance).

\citet{JCS06} presented spatially-resolved maps of PegDIG
at 4.5 and 8 $\mu$m obtained with the Spitzer IRAC camera. These
maps show only low surface brightness emission from the galaxy,
indicating low amounts of hot dust grains and PAHs in the ISM.
This, \citet{JCS06} propose, is the result of the destruction of grains by
supernova shocks and the inability of the ISM to regrow them in a
regime of low or zero star formation rates.


\section{The SDSS data for the Pegasus dwarf irregular galaxy}
\label{txt:SDSS_data}

An image of PegDIG was created by extracting the
relevant area from the SDSS data release DR5 \citep{DR5} imaging data set.
Owing to the distance of PegDIG, the high stellar densities in
its central regions, and an average seeing of $\sim 1.5''$ in
the SDSS imaging data, individual stars can no longer be resolved
as point sources in this area.
We performed surface photometry separately for the inner galaxy parts,
and for the outer parts, where stars still can be resolved.
The two independently-derived surface photometry results were subsequently
combined into a single, global surface photometry profile using
areas of overlap.  The results for the outer part were matched
to those for the inner part, yielding a single surface brightness
profile (SBP) per band for the entire galaxy.

\begin{table}
    \begin{center}
    \caption{Total magnitudes of the Pegasus dwarf irregular galaxy
    \label{tbl:Intergral1}}
    \begin{tabular}{ccccc} \hline
   \MC{1}{c}{Filter}
   & \MC{2}{c}{Circular Apertures}
   & \MC{1}{c}{2D GALFIT}
   & \MC{1}{c}{A$_\lambda$}
   \\ \hline
   & \MC{1}{c}{Apparent}
   & \MC{1}{c}{Total}
   & \MC{1}{c}{Total}
   &
   \\
   & \MC{1}{c}{[mag]}
   & \MC{1}{c}{[mag]}
   & \MC{1}{c}{[mag]}
   & \MC{1}{c}{[mag]}
   \\
    \MC{1}{c}{(1)}& \MC{1}{c}{(2)}  &\MC{1}{c}{(3)}  &\MC{1}{c}{(4)} &\MC{1}{c}{(5)}\\ \hline
    $u$           &  13.64$\pm$0.03 & 13.60$\pm$0.46 &13.20$\pm$0.03 &0.34          \\             
    $g$           &  12.26$\pm$0.02 & 12.23$\pm$0.20 &12.01$\pm$0.02 &0.25          \\             
    $r$           &  11.67$\pm$0.02 & 11.63$\pm$0.17 &11.48$\pm$0.02 &0.18          \\             
    $i$           &  11.35$\pm$0.02 & 11.32$\pm$0.21 &11.16$\pm$0.02 &0.14          \\             
    $z$           &  11.37$\pm$0.03 & 11.32$\pm$0.29 &11.08$\pm$0.03 &0.10          \\             
    \hline
    $(u-g)$       &                 & ~~1.37$\pm$0.50&~1.19$\pm$0.04 &              \\
    $(g-r)$       &                 & ~~0.60$\pm$0.26&~0.53$\pm$0.03 &              \\
    $(r-i)$       &                 & ~~0.31$\pm$0.27&~0.32$\pm$0.03 &              \\
    $(i-z)$       &                 & ~~0.00$\pm$0.36&~0.08$\pm$0.04 &              \\ \hline
     $U$          &                 &   12.88        &12.49$\pm$0.05 &  0.36        \\
     $B$          &                 &   12.68        &12.43$\pm$0.03 &  0.28        \\
     $V$          &                 &   11.87        &11.69$\pm$0.03 &  0.22        \\
     $R_c$        &                 &   11.41        &11.27$\pm$0.04 &  0.18        \\
     $I_c$        &                 &   10.89        &10.74$\pm$0.05 &  0.13        \\
    \hline
%
%
%
    \end{tabular}
    \end{center}
\end{table}

\subsection{Integrated photometry and Surface Brightness Profiles}
\label{txt:Integrated}

PegDIG itself was not identified as a separate entity by the
standard SDSS pipeline (see, for example, the SDSS DR6 database).
The integrated photometry, the creation of
SBPs, and their analysis were done in the manner described in
detail by \citet{Kniazev04}. First, $g$, $r$ and $i$ images for a wide
region around the galaxy were extracted from the SDSS database.
These were combined with weights to form an image of the object that is
deeper than any of the single SDSS bands could offer and that has
essentially the same stellar point spread function
(FWHM$\simeq 1''$) as the pipeline-reduced SDSS data.
The central part of the combined $gri$ image is shown in
Figure~\ref{fig:Direct_image}.
The galaxy location was defined above the 3$\sigma$ noise level on
the smoothed combined image, and all background sources were
subtracted using SDSS database coordinates and using additional
masking. The background was
subtracted and SBPs for all $ugriz$ filters were created in
circular apertures with a uniform isophote step size of 2\arcsec. The calculated
profiles were fitted on a logarithmic scale, following
\citet{Sers68}:
\begin{equation}
\label{equ:exp_profile}
\mu(r) = \mu_0 + 1.086 \cdot (r/\alpha)^n,
\end{equation}
where $R$ is the distance along the axis and $n$ is an
additional parameter.

To check the stability of our results we performed additional
two-dimensional modelling with the GALFIT program \citep{peng02}
using a one-component S\'ersic function for all $ugriz$ filters.
To exclude all foreground stars from the fitting procedure we used
the same mask image that was calculated on the previous step, where the
method of circular apertures was used.

The results for both methods are summarised in Tables~\ref{tbl:Intergral}
and ~\ref{tbl:Intergral1}.
Table~\ref{tbl:Intergral} presents all the model parameters for both methods,
except for the total magnitudes.
In case of circular apertures, the b/a axis ratio was calculated only once
when building the mask for the galaxy location and was kept constant
for the different filters. When GALFIT was used, the b/a axis ratio was allowed
to vary as one of the parameters of the fitted model.
GALFIT uses the following form for the S\'ersic function:
\begin{equation}
\label{equ:exp_profile1}
\mu(r)=\mu_e  + \left\{-b_n \left[\left({\frac{r}{r_e}}\right)^{n} -  1\right]\right\}
\end{equation}
where $\mu_e$ is the effective surface brightness and $r_e$ is the effective (half-light)
radius. Generally $\alpha = r_e/b_n$, but $b_n$ is different for each $n$ and
can be only found numerically.
A good approximation that can be used is $b_n \sim 2/n - 0.324$, but this is valid only
for $0.07\le n \le1$ \citep{TGC01} and is not correct here, since $n>1$.
For this reason, our recalculations
from $r_e$ to $\alpha$ are presented in brackets in column 7 ($r_e$) of Table~\ref{tbl:Intergral}.
The GALFIT solution for the $u$ filter was unstable and for this reason we
fixed $r_e$ to the same value it was for $g$. This parameter is marked
with a colon.

Table~\ref{tbl:Intergral1} presents the total magnitudes.
In column 2, the apparent magnitudes derived by integrating the luminosity
within a circular aperture out to the limiting isophotal level of the SDSS
data are listed.
The total magnitudes presented in Table~\ref{tbl:Intergral1}
were calculated extrapolating out to $r=\infty$
using equation (9) from \citet{Kniazev04} and the
parameters of the S\'ersic function listed in Table~\ref{tbl:Intergral}.
The rather large errors of these magnitudes result from the big errors of
the input parameters in Table~\ref{tbl:Intergral}.
The total magnitudes from GALFIT are also a result of flux integration
out to $r=\infty$. In the case of GALFIT, the total magnitude is a free parameter
of the fit.
We also presented $UBVR_cI_c$ magnitudes that were recalculated
from the $ugriz$ magnitudes using of equations from \citet{SDSS_phot1}.
The magnitudes in Table~\ref{tbl:Intergral1} are not corrected for foreground
Milky Way extinction, but extinction corrections calculated using the \citet{Schlegel98}
prescription are shown in column 5 of this table.

The photometric parameters calculated with either of the methods
are very similar, but the total magnitudes resulting from GALFIT are systematically
larger. Comparing with the integrated photometry of \citet{Zee00} we note that our
2D GALFIT values for B$_T$, $(B-V)$ and $(U-B)$ are very similar to those of van Zee.

\subsection{The Resolved Outer Part}
\label{txt:Resolved}

\subsubsection{Photometric selection of PegDIG stars}
\label{txt:Selection}

Tracing out PegDIG to much fainter surface brightness
levels than allowed by the direct surface photometry described in
the previous section is based on selecting stellar objects that
probably belong to the galaxy and on rejecting those that likely belong to the
foreground. This method of empirical photometric filtering was first
described and used by \citet{Grill95}. It was first implemented for
SDSS data by \citet{Odenk01a} for the globular cluster Pal~5
and was used by \citet{Odenk01b} to
study the Draco dwarf spheroidal galaxy.  Variants
of this photometric filtering of resolved stellar point sources
in the SDSS were also developed and used by \citet{Rock02}, \citet{SZB07}, and
\citet{Coleman07}.

\begin{figure}
    \begin{center}
    \includegraphics[angle=-90,width=8.5cm,clip=]{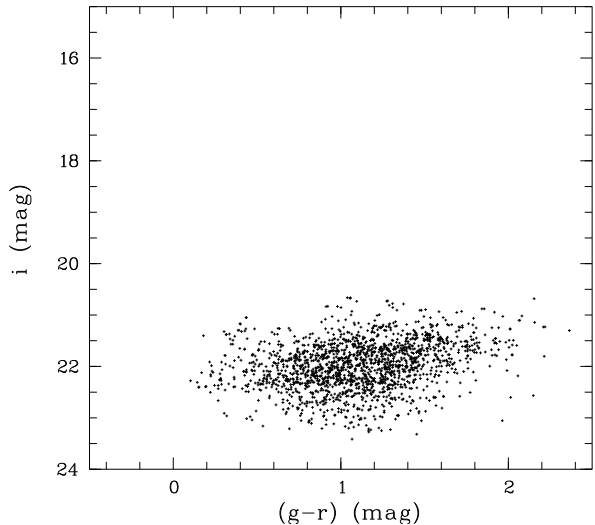}
    \caption{$i$ versus $(g-r)$ colour-magnitude diagram for stars
     selected
     in an elliptical annulus with a semi-major axis of 0.05--0.15 deg,
     ellipticity 0.6 and PA=$-55^\circ$,
     centred on ($\alpha,\delta$)=(352.15$^\circ$,+14.743$^\circ$).
    \label{fig:Pegasus_CMD}}
    \end{center}
\end{figure}

\begin{figure*}
    \begin{center}
    \includegraphics[angle=-90,width=5.7cm,clip=,bb=45 65 565 670]{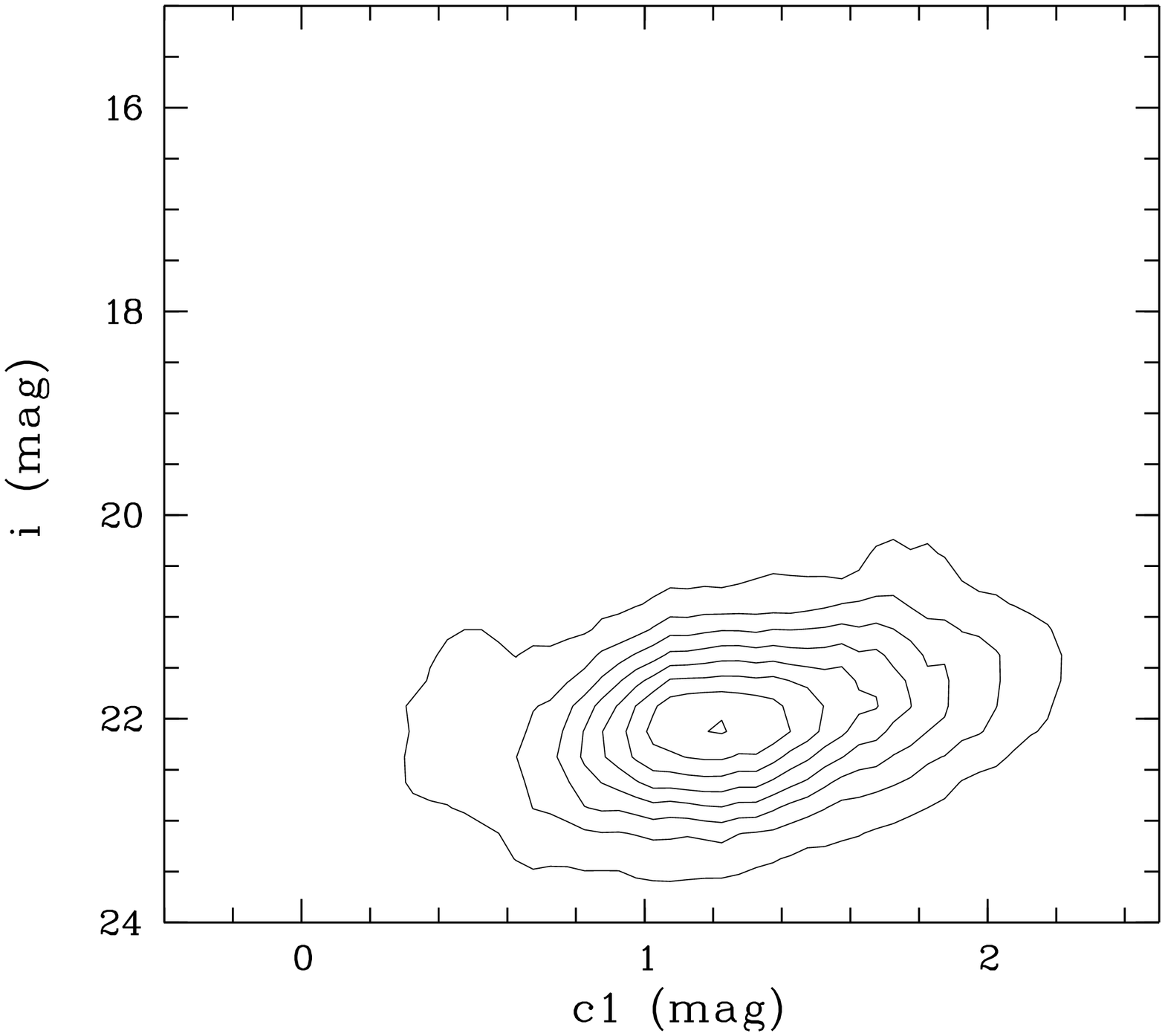}
    \includegraphics[angle=-90,width=5.7cm,clip=,bb=45 65 565 670]{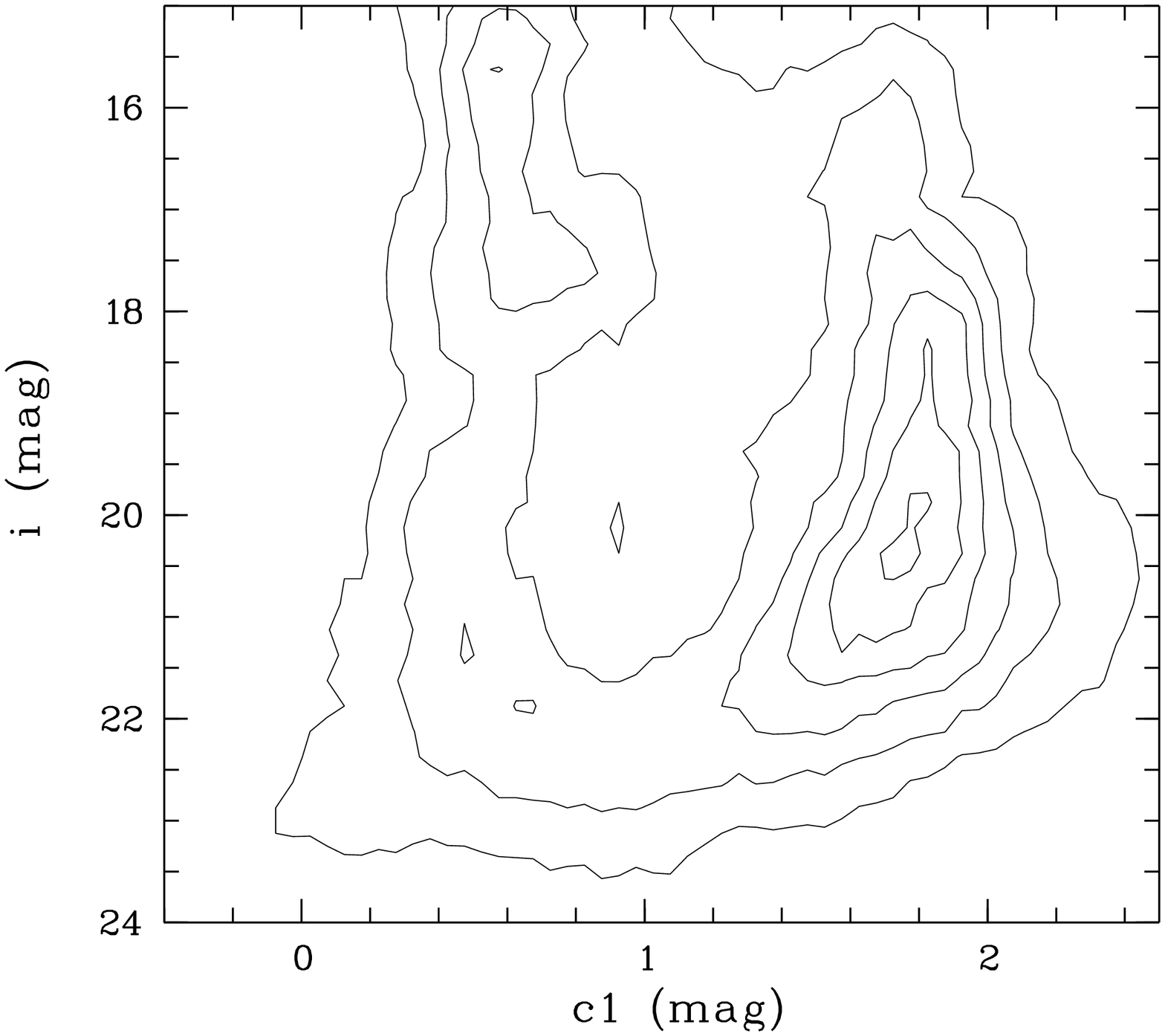}
    \includegraphics[angle=-90,width=5.7cm,clip=,bb=45 65 565 670]{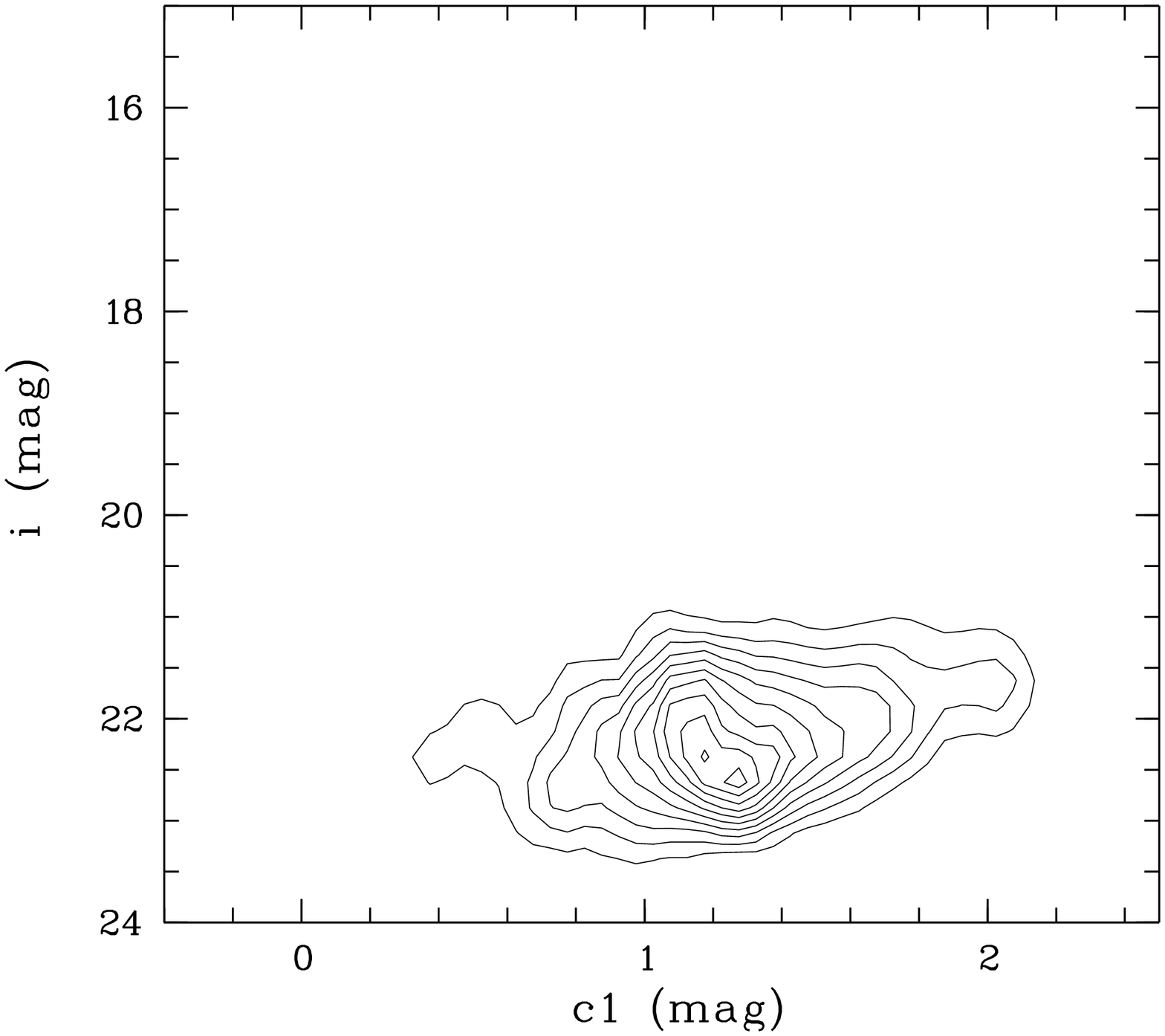}
    \caption{Separation of foreground Milky Way stars from the
    stars in PegDIG.
{\it Left:} Density distribution $f_P$ of Pegasus stars in the
$(c_1,i)$ colour-magnitude plane that is shown as contour plot.
Stars were taken from an elliptical annulus as described for
Fig.~\ref{fig:Pegasus_CMD}.
{\it Middle:} Density distribution $f_F$ of field stars in the
$(c1,i)$ colour-magnitude plane. Stars were calculated outside of
an ellipse with similar parameters and with a semi-major axis of 0.25 deg.
{\it Right:} Lines of constant number ratio (the population contrast)
$s = f_P / f_F$.
    \label{fig:Dens_distr}}
    \end{center}
\end{figure*}

\begin{figure}
    \begin{center}
    \includegraphics[angle=-90,width=7.9cm,clip=]{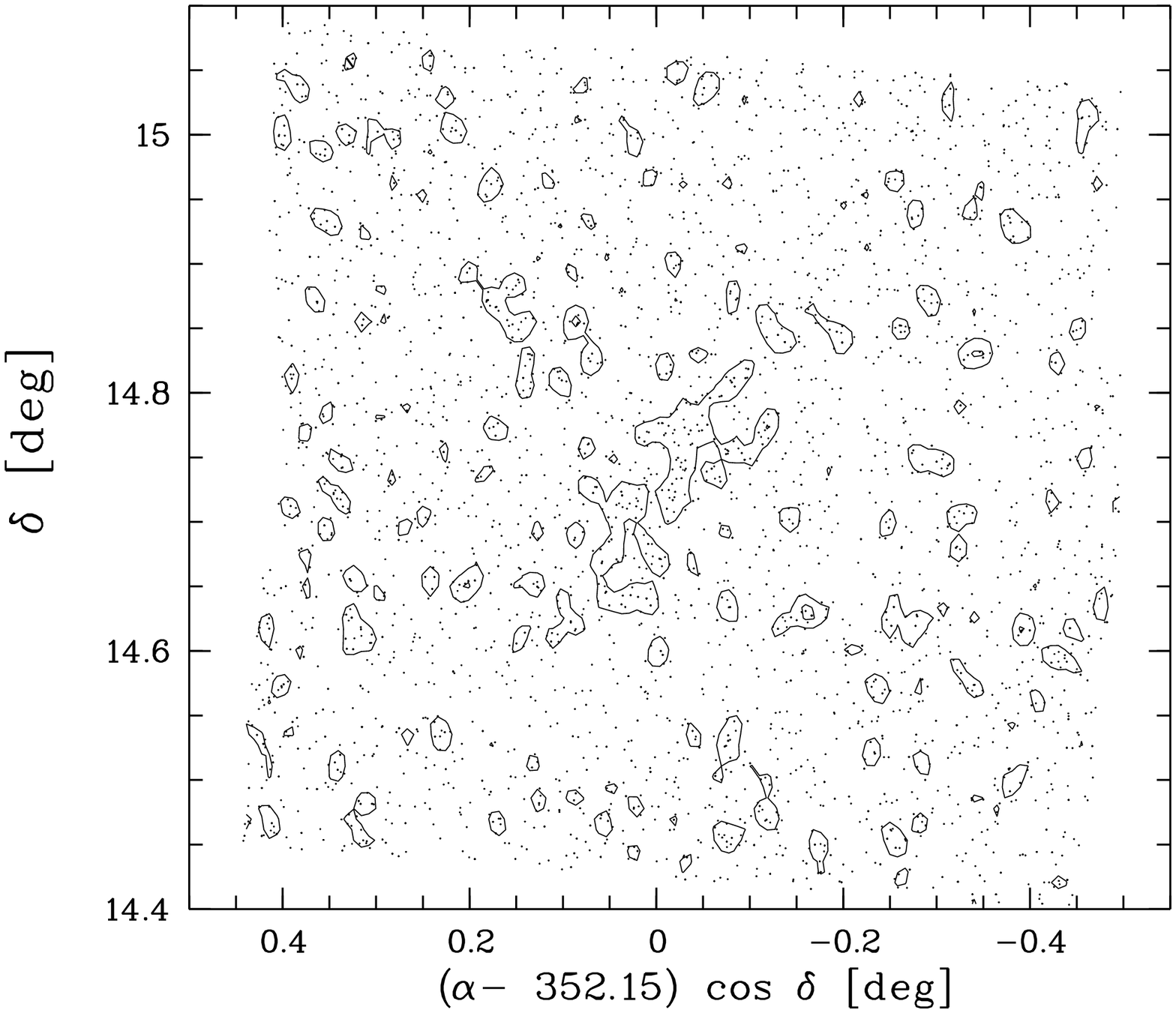}
    \includegraphics[angle=-90,width=8.0cm,clip=]{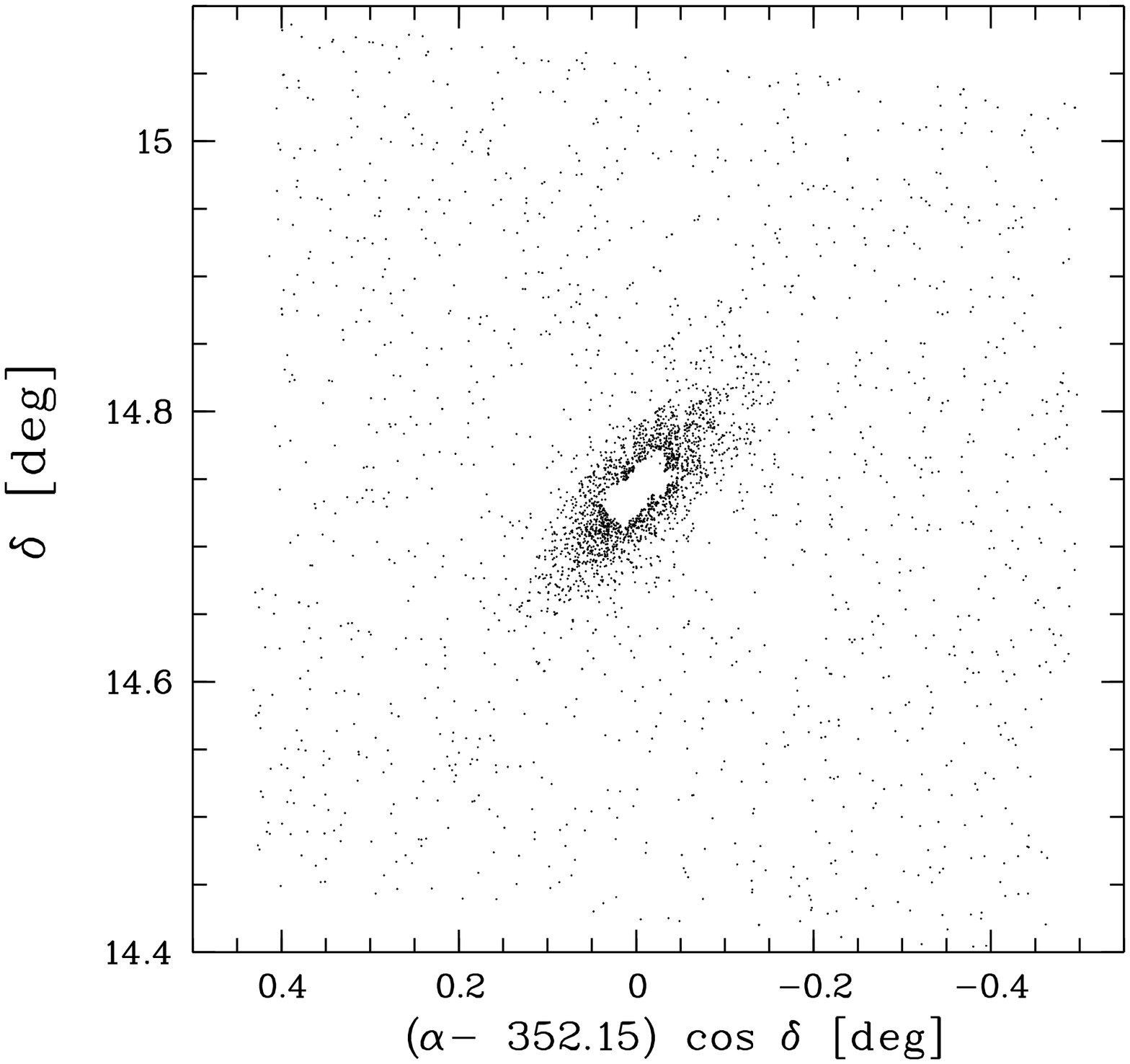}
    \caption{
    The spatial distribution of SDSS selected point sources
    in the direction of the Pegasus dwarf galaxy.
    {\it Top panel:}
    Stars selected by the algorithm as belonging to the Galactic foreground
    appear randomly distributed in this image.
    Contours of equal stellar surface density at a level of 1$\sigma$
    and 2$\sigma$ are shown with thin lines ($\sigma$ is the rms background variation).
    There are some concentrations at only a 1$\sigma$ level in the center.
    {\it Bottom panel:} Final spatial distribution of $\sim$3400 photometrically
    selected stars in the PegDIG area. The stars selected as candidate PegDIG
    members are concentrated around the recognised location of the galaxy.
    \label{fig:Pegasus_point_cleaned}}
    \end{center}
\end{figure}

First, we selected all point sources imaged by the survey in a $\sim$0.6 square
degree region centred on the Pegasus dwarf galaxy.
Approximately 8000 point sources
in this area were classified by the standard SDSS pipeline as
stars. Since the SDSS detects only the most luminous stars in
PegDIG, the number of identifiable
stars belonging to the galaxy is quite limited. For this reason we
did not use any additional selection criteria, but rejected only
the sources for which the photometric error in any of the three
most sensitive passbands ($gri$) was larger than 0.4 mag.
The spatial distribution of all the selected sources
has a mean stellar
density of $\sim$2.85 stars arcmin$^{-2}$ for the foreground stars
(all areas, excluding the central part of the field where PegDIG
is obviously located).

The crucial point in constructing a high-contrast map for the outer
part of the Pegasus system is to achieve an efficient
discrimination between the tracers of this system and the
foreground field stars. Following \citet{Odenk01b},
we created a training set by selecting $\sim$1500 stars
from the central part of the
PegDIG (defined as an ellipse with a semi-major axis of
0$^{\circ}$.15, ellipticity 0.6 and PA=$-55^\circ$, centred on
($\alpha, \delta$)=(352.15$^\circ$,+14.743$^\circ$).

In Fig.~\ref{fig:Pegasus_CMD} we show a colour-magnitude diagram of the
point sources that are candidate members of PegDIG as mentioned above.
The diagram shows a fairly amorphous, extended plume of faint
sources that resemble the distribution of stars in other distant
dIrr galaxies from early CCD observations \citep[e.g,][]{Tosi91,Greggio93}.
The SDSS data share several characteristics with these early, ground-based
CCD data:  For a galaxy at the distance of PegDIG, they are close to the
detection limit and they are strongly confusion and crowding limited.
The extended plume of stars in the SDSS stellar point source catalog
is composed of young main-sequence stars, blue loop stars, red supergiants,
luminous asymptotic giant branch stars, and stars at the tip of the
red giant branch.  In other words, we are sampling the most luminous stars
in PegDIG, which represent a range of different stellar populations and
ages covering many billions of years.  While these shallow data are not
suited nor intended for detailed studies of the star formation history
of PegDIG, they are very valuable for studies of its structure.  The
fact that stars on the upper red giant branch are included is particularly
important, since this will be helpful when trying to trace the faint outer
limits of PegDIG, which are typically dominated by old(er) populations.

Following the procedure, described by \citet{Odenk01a, Odenk01b},
we calculated new colour indices  $c_1$ and $c_2$:

\begin{equation}
\label{eq.PCA}
\begin{array}{llr}
c_1 & = & 0.921 \cdot (g-r) + 0.389 \cdot (r-i)  \\ [0.1cm]
c_2 & = & -0.389 \cdot (g-r) + 0.921 \cdot (r-i)
\end{array}
\end{equation}

The relations in the equations~\ref{eq.PCA} are very close to
those derived by \citet{Odenk01b} for the Draco dSph
and similar to those derived by \citet{Odenk01a} for Palomar~5.

After constructing the colour index $c_1$, which is a linear
combination of the $(g-r)$ and $(r-i)$ colours, the next step was
to design an empirical filter mask in the $(c_1,i)$
colour-magnitude (CM) plane that will allow the optimal separation of the
colour-magnitude distribution of sources belonging to the PegDIG
population from that of the foreground field stars.
The basic assumption is that the stars in PegDIG are of
specific types, based on their colours and apparent magnitudes, and
will occupy a specific locus in the CM diagram.
This is shown in Figure~\ref{fig:Dens_distr}.
The region with maximal $s$ values in the right panel of
Figure~\ref{fig:Dens_distr} delineates the locus in the CM
plane with the highest fraction of PegDIG stars relative to field
stars. This last figure indicates that the stars selected from
the SDSS data set as probable PegDIG members are mostly between
21$\leq i \leq$23.5 mag.

The next step was to identify a level $s_{opt}$ that yields the
highest contrast for the selected area of PegDIG compared to the field
stars. This implies that a filter mask for the selection of
Pegasus stars should include as many points as possible with $s
\ge s_{opt}$, where $s_{opt}$ is a threshold value. To find the
optimal number density threshold $s_{opt}$ we computed the
signal-to-noise ratio (SNR) for a range of $s$ values, using the
following equation from \citet{Grill95} and from \citet{Odenk01a,Odenk01b}:
\begin{equation}
{\rm SNR(s)} = \frac{N_P(s) - w N_F(s)}{\sqrt{N_P(s) + w^2 N_F(s)}},
\end{equation}
Here $N_P(s)$ is the total number of stars in the sample defined
by $s$ and for the region of PegDIG, and $N_F(s)$ is the number of
stars in the same sample, but for the region where the foreground
population is probed. The $w$ parameter scales areas of
these two regions. Since we are interested in the outer part of
PegDIG, the SNR was optimised first for the elliptical
annulus with a semi-major axis between 0.12 and 0.20 deg from the
centre of Pegasus. After that, other areas and different
foreground regions were tested; we found  that $s_{opt}$ was practically
identical in all those tests.

The final selection filter constructed here removed $\sim$76\% of
the contaminating field stars and reduced the mean density of the
contaminating foreground stars from 2.85 stars arcmin$^{-2}$ to $0.69\pm1.67$
stars arcmin$^{-2}$. The spatial distribution of stars
identified as foreground sources is shown in the top panel of
Figure~\ref{fig:Pegasus_point_cleaned}. It is clear that these stars are
randomly distributed and do not show any strong concentration
toward the bright core of PegDIG; a concentration could be
expected if the filtering operation had been inefficient.
The spatial distribution of stars with characteristics matching
our final selection filter is shown in the bottom panel of
Figure~\ref{fig:Pegasus_point_cleaned}. A visual comparison of the
top and bottom panels of this figure shows that the objects
selected by the filter as candidate PegDIG members do concentrate
around the known galaxy, implying that the filtering operation
indeed selected preferentially PegDIG stars very distant from the
unresolved inner body.

\begin{figure}
    \begin{center}
    \includegraphics[angle=-90,width=8.0cm,clip=]{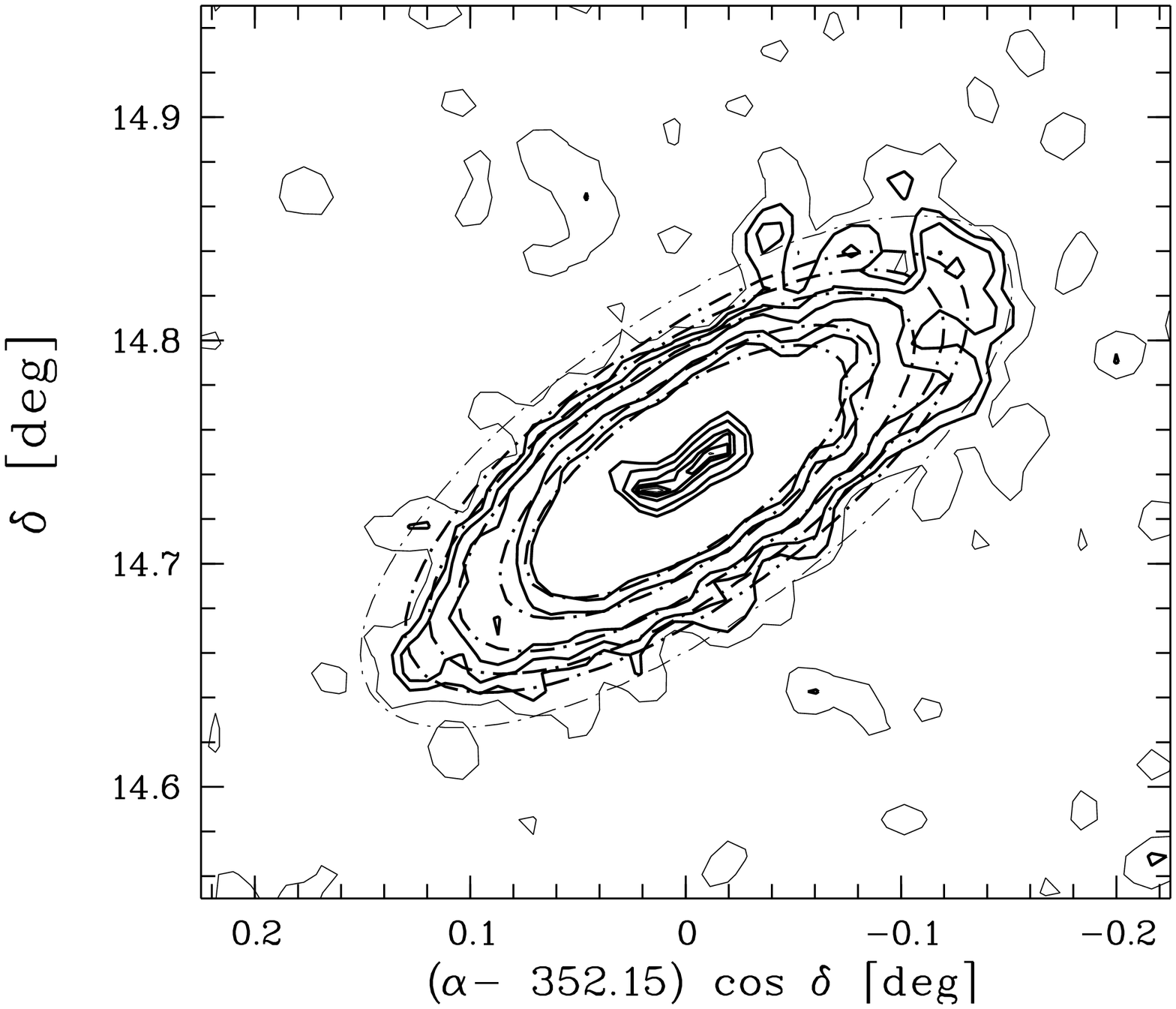}
    \includegraphics[angle=-90,width=8.0cm,clip=]{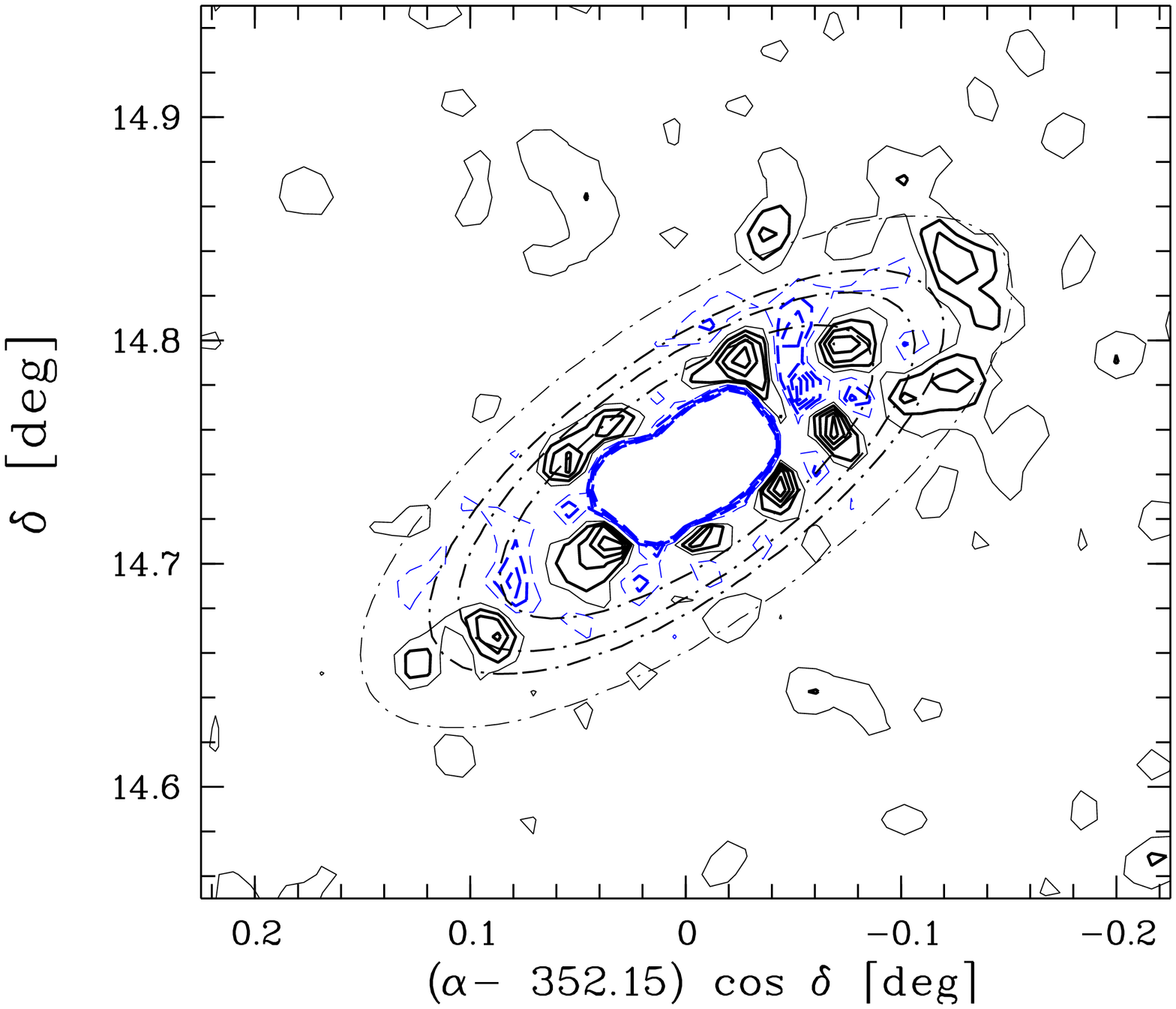}
    \caption{{\it Top:}
    Contour plots of the observed distribution of stars in PegDIG.
    Contours of equal stellar surface density are shown
    at 2$\sigma$, 3$\sigma$, 5$\sigma$ and 10$\sigma$
    with a thinner contour at 1$\sigma$.
    The profiles in the centre of the galaxy show
    a ``hole'' produced by missing information in the high stellar
    density part of the galaxy.
    Dot-dashed lines show the contours of the best-fit exponential model
    (2D S\'ersic profile) at $1\sigma$, $3\sigma$, $5\sigma$ and $10\sigma$ levels.
    Note the ``dendritic'' extensions of the faint contours in the
    north-west part of the galaxy.
    {\it Bottom:} Residuals of the fit shown in the top panel, rescaled to the level
    of the mean background counts. As before, the contour with thinner lines
    corresponds to levels of $\pm 1\sigma$.
    Short-dashed (blue) lines show negative levels of the residuals.
    \label{fig:Pegasus_cleaned_cent}}
    \end{center}
\end{figure}

A stellar number density map of the central part of the studied field
is shown in Figure~\ref{fig:Pegasus_cleaned_cent} as isopleths.
The surface density was derived through counts on a
30\arcsec\ by 30\arcsec\ grid and subsequent weighted averaging
within a radius of one grid step. The thin lines show contours at
the 1$\sigma$ level above the mean background density, where
$\sigma$ is the rms of the background stellar density
fluctuations, showing the lowest level at which the PegDIG stars
start to be recognisable.
Since any significant detection of the PegDIG population requires a
surface density of at least 2$\sigma$ above the background, this
level is marked by the thick contour in
Figure~\ref{fig:Pegasus_cleaned_cent}. Levels of
3$\sigma$, 5$\sigma$ and 10$\sigma$ are plotted there with thick
lines as well, to reveal the shape of the galaxy at different
stellar number densities; the plot shows that the overall distribution
of PegDIG stars is approximately ellipsoidal, but that the outer
contours appear deformed. It is also obvious that the NW end of the galaxy
seems to be much more irregular than the SE one. These
deviations, at the NW end, are very strong, showing up even
at the 3$\sigma$ contour.

\begin{table*}
\begin{center}
\caption{Parameter values for the best-fit model for the surface density distribution of PegDIG
\label{tab:2D-fit}}
\begin{tabular}{ccccccc} \hline
\MC{1}{c}{Model}& \MC{1}{c}{$\alpha_c$} & \MC{1}{c}{$\delta_c$} &\MC{1}{c}{b/a}   &\MC{1}{c}{PA}        & \MC{1}{c}{$\alpha$}   &\MC{1}{c}{$n$} \\
	 & \MC{1}{c}{[2000.0]}   & \MC{1}{c}{[2000.0]}   &               &\MC{1}{c}{[degree]}  & \MC{1}{c}{[arcsec]}&               \\
\hline
2D S\'{e}rsic       & 23:28:35.52 &+14:44:29.04 & 0.39$\pm$0.01 &$-$55\fdg7$\pm$0\fdg1&   138$\pm$2     & 1.11$\pm$0.02 \\
1D S\'{e}rsic       &             &             & 0.40          &$-$55\fdg7           &   192$\pm$35    & 1.36$\pm$0.21 \\
\hline \hline
\MC{1}{c}{Model}& \MC{1}{c}{$\alpha_c$} & \MC{1}{c}{$\delta_c$} &\MC{1}{c}{b/a}   &\MC{1}{c}{PA}        & \MC{1}{c}{$r_c$}   &\MC{1}{c}{$r_t$} \\
\hline
2D King 62       & 23:28:35.02 &+14:44:28.30 & 0.38$\pm$0.03 &$-$55\fdg9$\pm$0\fdg1& 39$\pm$20               &    957$\pm$50           \\
\hline
\end{tabular}
\end{center}
\end{table*}

\begin{figure}
    \begin{center}
    \includegraphics[angle=-90,width=8.0cm,clip=]{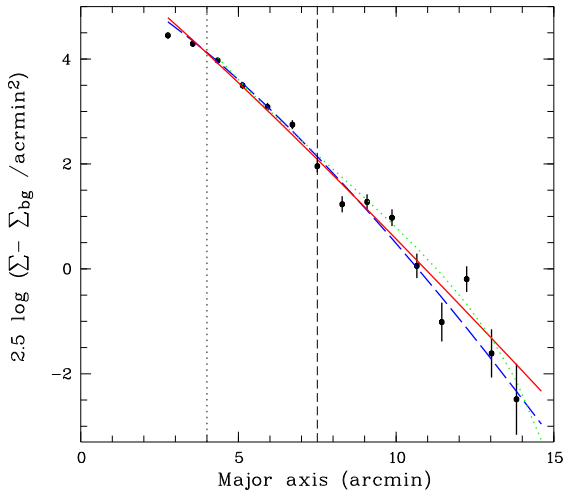}
    \caption{Radial profile of the surface density of stars in PegDIG.
    Data points with error bars show the profile of the observed distribution
    (star counts in elliptical rings around the centre of PegDIG with
    parameters of the best-fit model shown in
    Table~\ref{tab:2D-fit},  and with the mean background
    0.69 stars arcmin$^{-2}$ subtracted).
    The red solid line shows the major axis model distribution for the
    GALFIT 2D S\'ersic model (see Table~\ref{tab:2D-fit} for parameters).
    The blue long-dashed line shows the best-fit 1D S\'ersic model, where
    the point at $\sim$12 arcmin was excluded since it represents the extensions
    visible in the bottom panel of Figure~\ref{fig:Pegasus_cleaned_cent}.
    When doing a fit with this point included, the result of the 1D fitting
    is very close to the result of 2D with GALFIT.
    The green dotted line shows the major axis model distribution for the
    GALFIT 2D King model (see Table~\ref{tab:2D-fit} for parameters).
    Points located to the left of the vertical dotted line
    at R=4 arcmin were not included in the fit for either model.
    The vertical short-dashed line indicates a radial distance of $\sim$450 arcsec,
    where our SBP profiles in the $ugriz$ filters terminate.
    \label{fig:Pegasus_dens_profile}}
    \end{center}
\end{figure}

\subsubsection{Structure of PegDIG}
\label{txt:Peg_2Dfit}

To quantify the size, shape and orientation of the Pegasus dwarf
irregular galaxy we fitted a two-dimensional model to the observed surface
density distribution. For that, the surface density was sampled on
a 30\arcsec\ $\times$ 30\arcsec\ grid of non-overlapping cells. We
performed two-dimensional modelling using GALFIT
with a one-component S\'ersic function. The central
part of the galaxy (which looks like a hole in our data) was
excluded from the modelling with the special mask allowed by GALFIT.
The best-fit parameters are given in Table~\ref{tab:2D-fit}
and the best fit is shown in
Figures~\ref{fig:Pegasus_cleaned_cent} and
\ref{fig:Pegasus_dens_profile}. The lower panel of
Figure~\ref{fig:Pegasus_cleaned_cent} shows the difference between
the actual data and the fitted model.

Although the statistical significance of any single feature may
not be extremely high, we note that there seems to be a ring-like
distribution of peaks (clumps) around the unresolved part of the galaxy.
The peaks in the ring configuration encircling the inner, unresolved
part of the galaxy are highly significant, more than 5$\sigma$ above
the background. The three peaks revealed to the north-west of the galaxy,
between the 1$\sigma$ and 3$\sigma$ contours of the model, are of possibly
even higher significance. These peaks could not be
just artefacts of Poisson noise, as suggested by \citet{M08} for other
faint nearby galaxies, but here they are sufficiently significant,
namely 3$\sigma$ or higher,
as shown in the lower panel of Fig.~\ref{fig:Pegasus_cleaned_cent}.
Such peaks are not seen at the opposite end
of the galaxy and represent deviations of the faint outskirts of
the galaxy from a pure S\'{e}rsic profile, which are restricted to the NW side.

As can be seen in Figure~\ref{fig:Pegasus_cleaned_cent}, the highest
density of detected stars in the area of the peaks (clumps) is about 10--20 $\sigma$
of the estimated background noise scatter. With the latter value
1.67 star arcmin$^{-2}$ (see Section~\ref{txt:Resolved}),
the 20 $\sigma$ level corresponds to a star density of 33.5 star arcmin$^{-2}$
or $\sim$0.009 star arcsec$^{-2}$. In other words, the area with the
highest density of detected stars in the SDSS Pegasus data corresponds
to about one star in a circle with a radius of $\sim$6 arcsec.
For an effective seeing of one arcsec and a pixel scale of 0.396 arcsec there
should not be any crowding problem in the regions with this and lower stellar density,
either for detection, or for object classification as a star/nebula.

To demonstrate the evidence for a much larger spatial extent of PegDIG
indicated by our data, we show in Figure~\ref{fig:Pegasus_dens_profile} the
observed radial stellar
density profiles. These were derived from star counts in elliptical annuli,
with parameters of the ellipses taken from Table~\ref{tab:2D-fit}.
The logarithm of the mean density above background is plotted
versus  the radius of each annulus. The radius refers to the distance
along the fitted major axis. The central $2.7'$ region was not included in
the  fit and is not shown.

The best-fitting generalised S\'{e}rsic model, described by the
parameters listed in Table~\ref{tab:2D-fit}, is characterised by a radial scale
length of $\alpha=138$ arcsec and an exponent $n=1.11$.
Comparing these numbers with those shown in Table~\ref{tbl:Intergral}
one can see that both the exponent $n$ of the density distribution
and the scale parameter $\alpha$ are very close
to the values derived from the $gri$ images.
As  Figure~\ref{fig:Pegasus_dens_profile} shows, the model
profile fits our data very well over the entire region;
there is no compelling evidence for a real cutoff in the radial
profile of PegDIG within the current observational limits.
Using SDSS data for the unresolved part of the galaxy we can
trace the galaxy only up to a $\sim$450\arcsec\ radius along the major
axis. Using star counts, and improving the contrast of our density
map, we can trace PegDIG up to $\sim$800\arcsec, but a distinct
edge of the galaxy still remains undetected.

The model fits provide a useful, smooth reference against which
to identify
peculiarities in the shape of the observed distribution of stars.
Such peculiarities are the above-mentioned ``ring'' of stellar
density peaks around the unresolved part of the galaxy, and the
three peaks at the NW end of the faint galaxy extension revealed
by our filtering procedure and subtraction of the fitted model. We
note that there are no counterparts to these peaks in the SE part
of PegDIG. These features will be discussed below.

\begin{figure}
    \begin{center}
    \includegraphics[angle=-90,width=8.0cm,clip=]{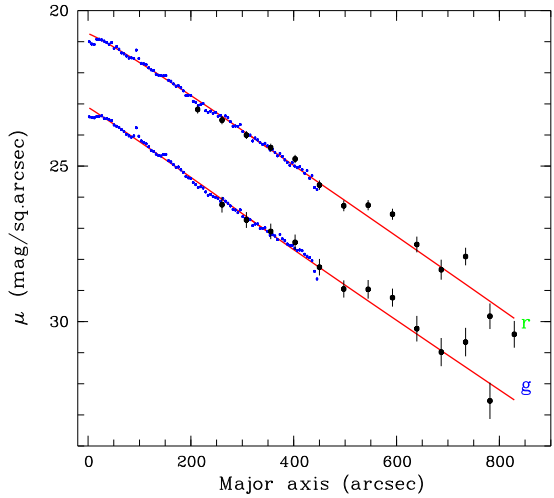}
    \caption{%
Composite SDSS $gr$-band SBPs.  The SBPs were calculated in circular apertures
for the unresolved part of the Pegasus dwarf irregular galaxy (blue points)
and recalculated to the major axis using an ellipticity of e=0.61.
The other data points for the SBP were calculated from star counts (black points) in
ellipses with parameters taken from the best-fit model for the stellar
surface density map and corrected for the part of the colour-magnitude diagram
that is not detected in the SDSS data. The red curves represent the best-fit
S\'ersic model calculated from the SBPs for the unresolved inner part of Pegasus
dwarf galaxy. The figures indicate that this model
also fits the outer, fainter parts of this object reasonobly well.
The SBP profile for the $r$ filter was shifted by an additional $-$2
magnitudes for clarity.
    \label{fig:SBP_tot_prof}}
    \end{center}
\end{figure}

\begin{figure}
    \begin{center}
    \includegraphics[angle=-90,width=8.0cm,clip=]{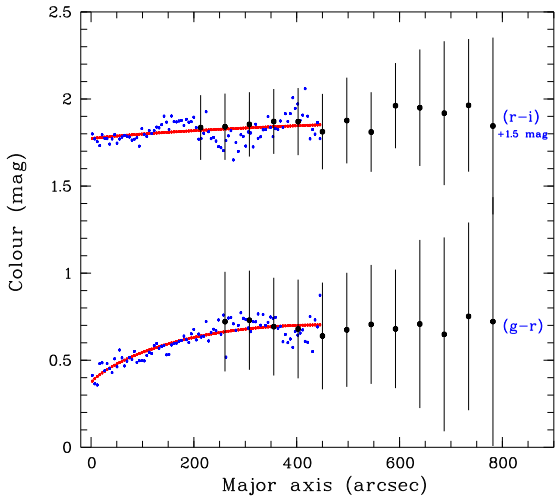}
    \caption{%
Composite SDSS $(g-r)$ and $(r-i)$ colour distributions.
One of the components of the colour distributions was calculated in circular
apertures for the unresolved part of PegDIG (blue points) and recalculated
to the major axis using an ellipticity e=0.61. The other component was calculated
from star counts (black points) in ellipses with parameters taken from the best-fit
to the stellar surface density map and corrected for the undetected part of
the colour-magnitude diagram. The red curves represent the best-fit S\'ersic models
that were calculated for the  SBPs created for the unresolved part of Pegasus dwarf galaxy.
The $(r-i)$ colour distribution was shifted by $+$1.5 magnitudes for clarity.
    \label{fig:SBP_tot_color}}
    \end{center}
\end{figure}

\subsubsection{Surface Brightness Profiles from Star Counts}
\label{txt:SBP_stars}

Our filtered star count data allow us to construct surface
brightness profiles (SBPs). Such SBPs can be derived using the
filtered star count data, summing stellar magnitudes in
elliptical annuli where the parameters of the ellipses are taken from
Table~\ref{tab:2D-fit}, and normalising them to the area of the
annuli.

Assuming that the PegDIG stellar populations do not vary wildly in the outer
parts of the galaxy (indeed they do not, as we show below), we can conclude
that one sees the same profiles regardless of the method used.
In the case of the unresolved part of PegDIG these were
calibrated correctly, but those constructed from the star counts
were not, since stars below the detection limit of the SDSS would
be excluded from the star counts. Using both data sets for the part
of the galaxy where the two methods overlap we can derive
factors for the different filters and derive composite SBPs down to the
faintest levels of PegDIG. These SBPs can be used to understand the
evolutionary history of the galaxy, since the star formation
in the outskirts of the galaxy presumably took place a long time ago.

The interpretation of these correction factors is that they compensate
the SBPs in the galaxy part created by using resolved PegDIG stars for
the missing light produced by the fainter stars that are not detected
on SDSS images. The underlying assumption is that the slope and cutoffs
of the IMF are constant between the inner unresolved part of PegDIG and
the outer part where the brighter stars are resolved and recorded.

\begin{figure}
\includegraphics[angle=0,width=7.0cm]{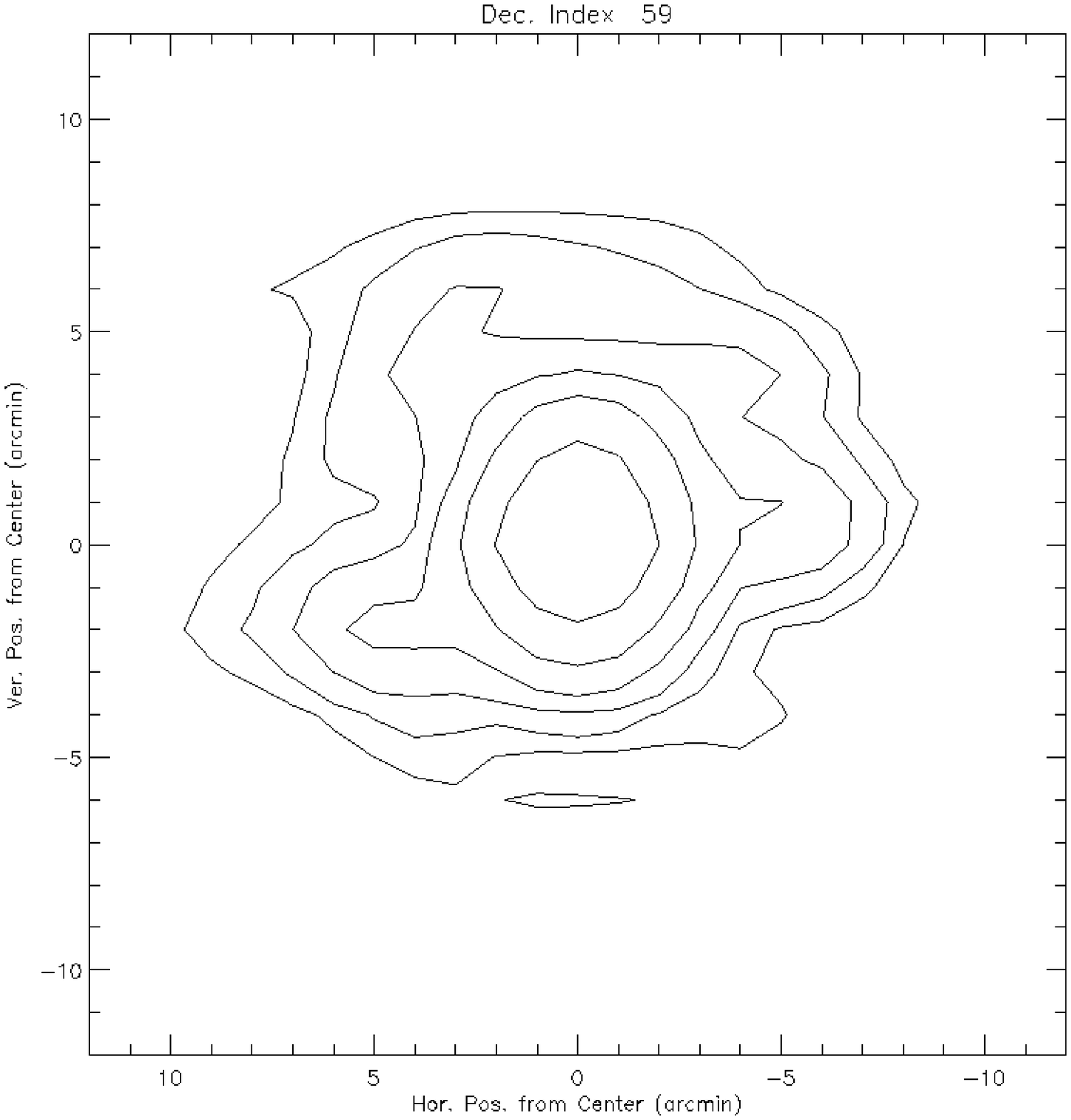}
\includegraphics[angle=0,width=7.0cm]{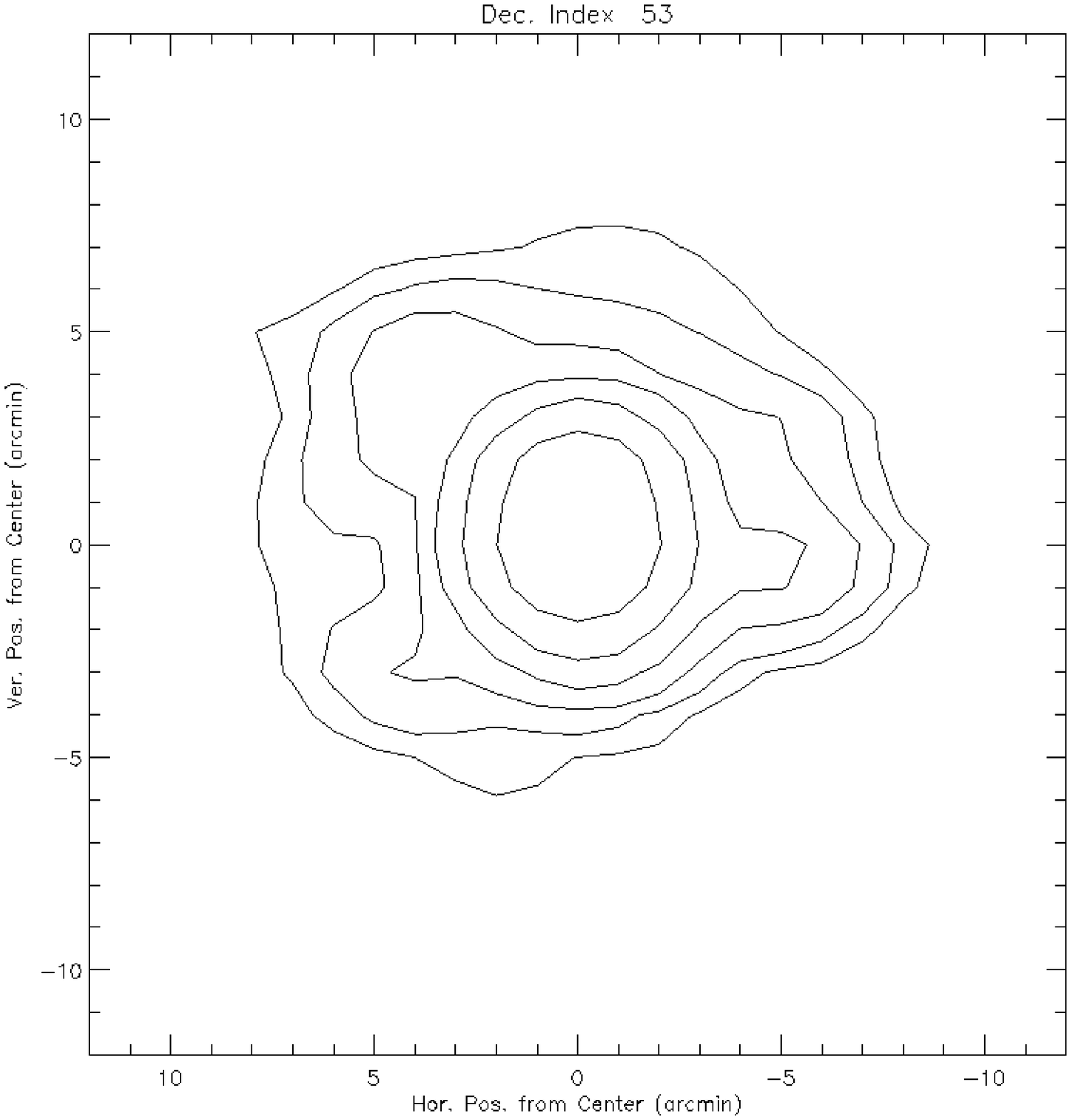}
\caption{Effective ALFALFA beams at two representative declinations.
Top shows declination $14\degr47^m30^s$ and bottom shows $14\degr41^m30^s$.
The contours are spaced logarithmically, in steps of 3.33 dB down from
the beam maximum.
\label{f:beams}}
\end{figure}

Composite SBPs in the $gri$ bands were constructed from the
unresolved part of the galaxy and from SBPs that were calculated
from the star counts. Some of these composite SBPs are shown in
Figure~\ref{fig:SBP_tot_prof}. Figure~\ref{fig:SBP_tot_color}
shows the composite $(g-r)$ and $(r-i)$ colours. Both the $(g-r)$
and $(r-i)$ composite colours show very stable values for the outer
parts of the galaxy, which justifies our previous statement.

\begin{figure*}
    \begin{center}
    \includegraphics[angle=0,width=4.5cm]{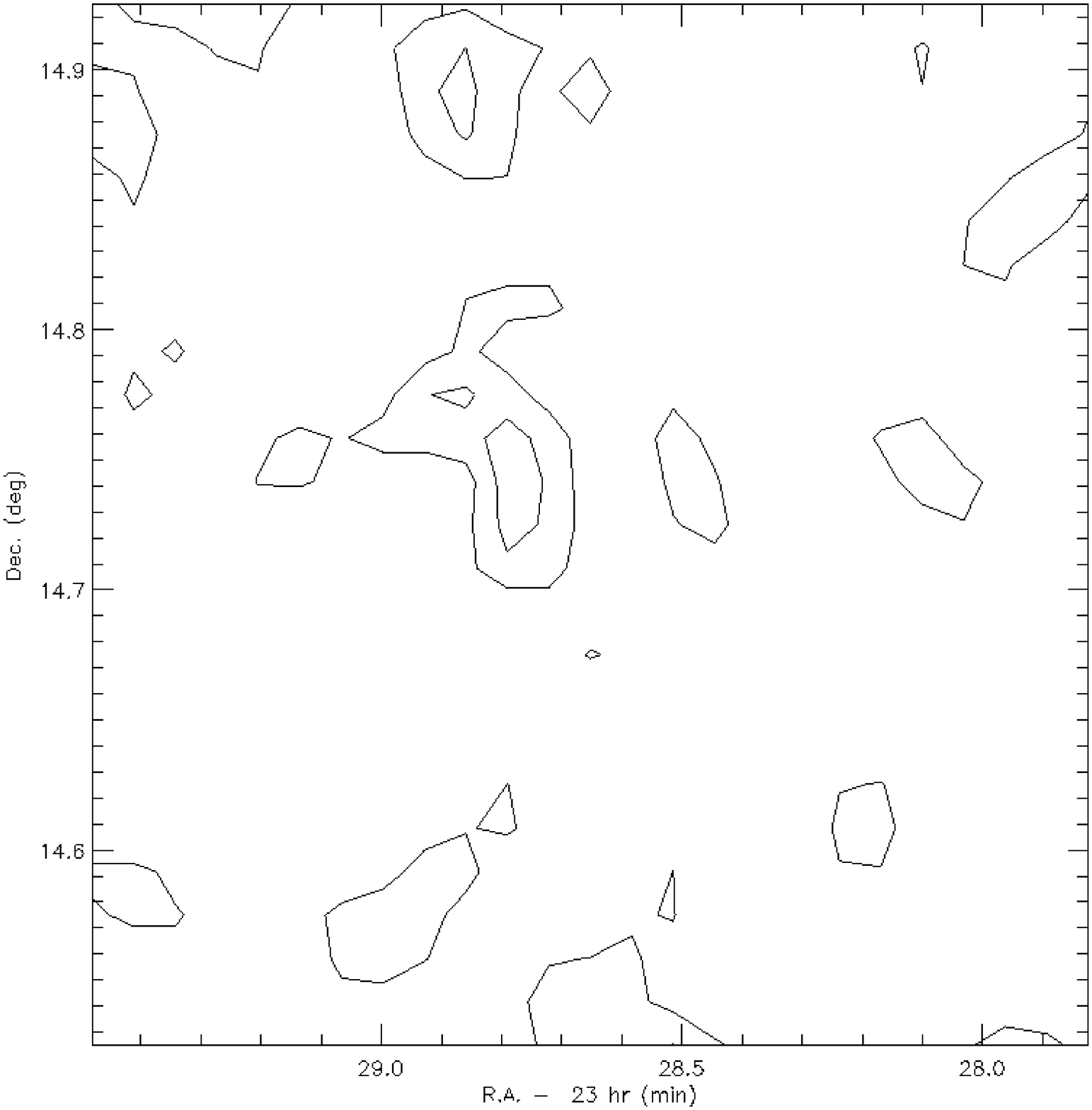}
    \includegraphics[angle=0,width=4.5cm]{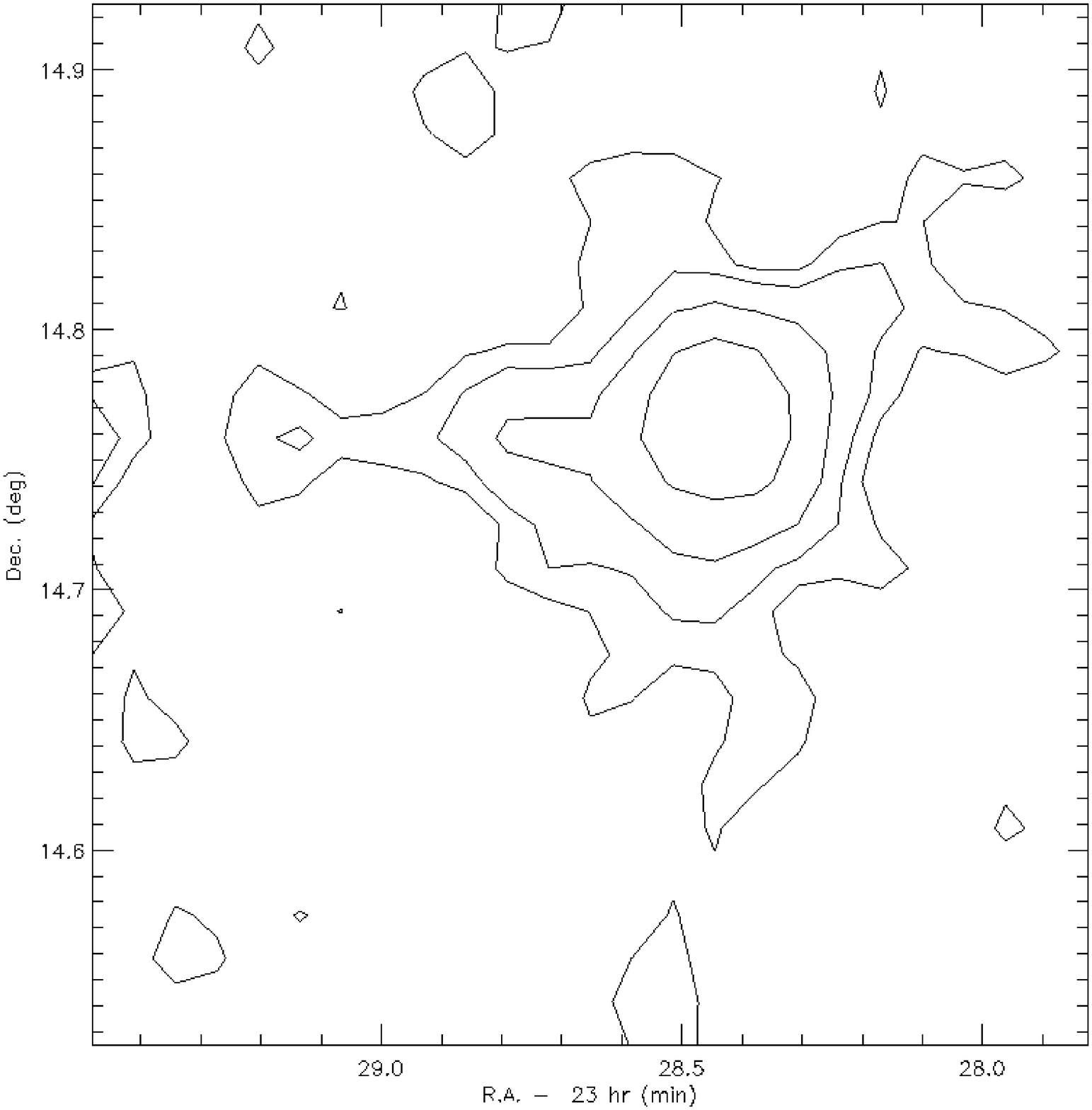}
    \includegraphics[angle=0,width=4.5cm]{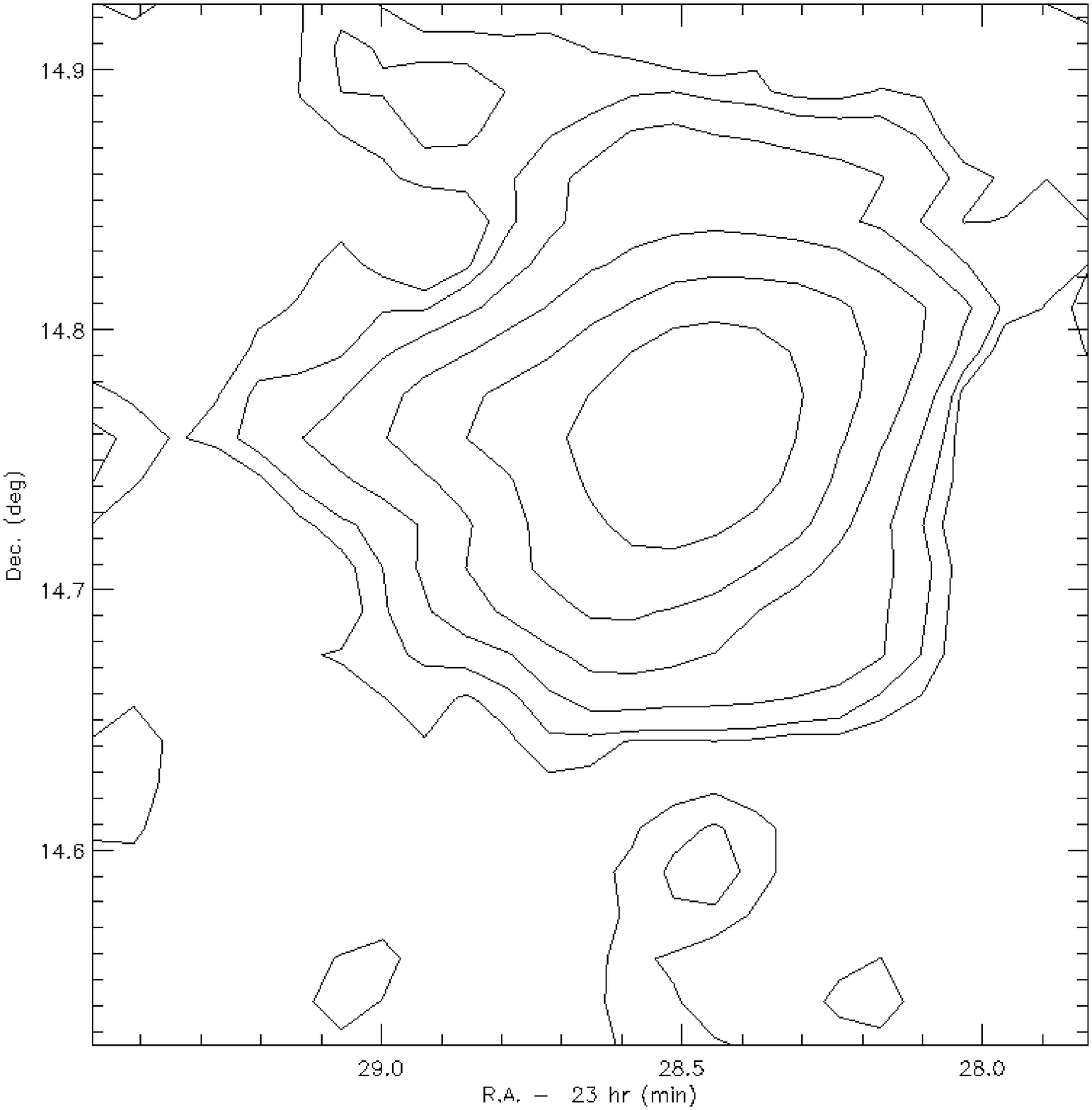}
    \includegraphics[angle=0,width=4.5cm]{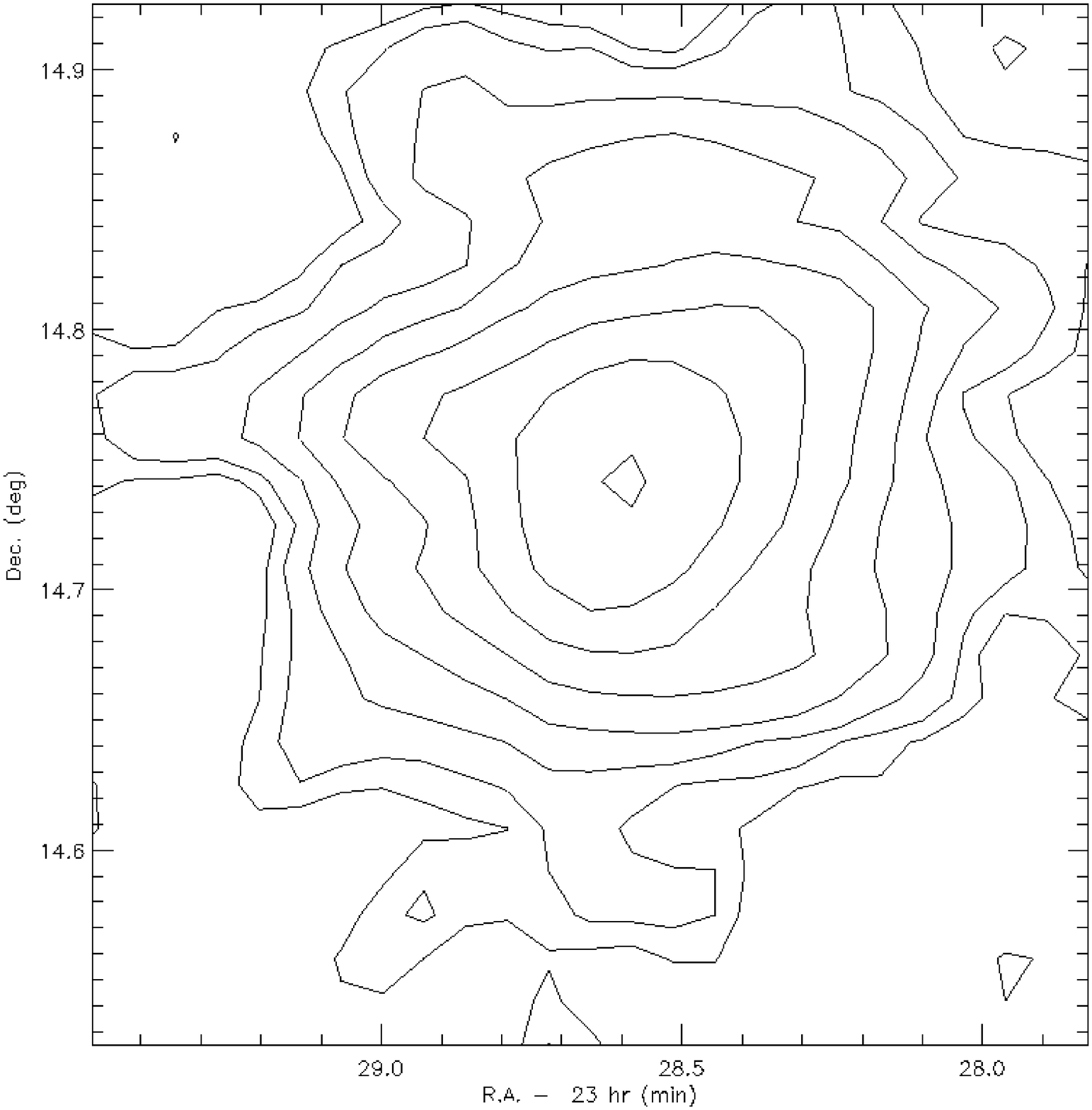}
    \includegraphics[angle=0,width=4.5cm]{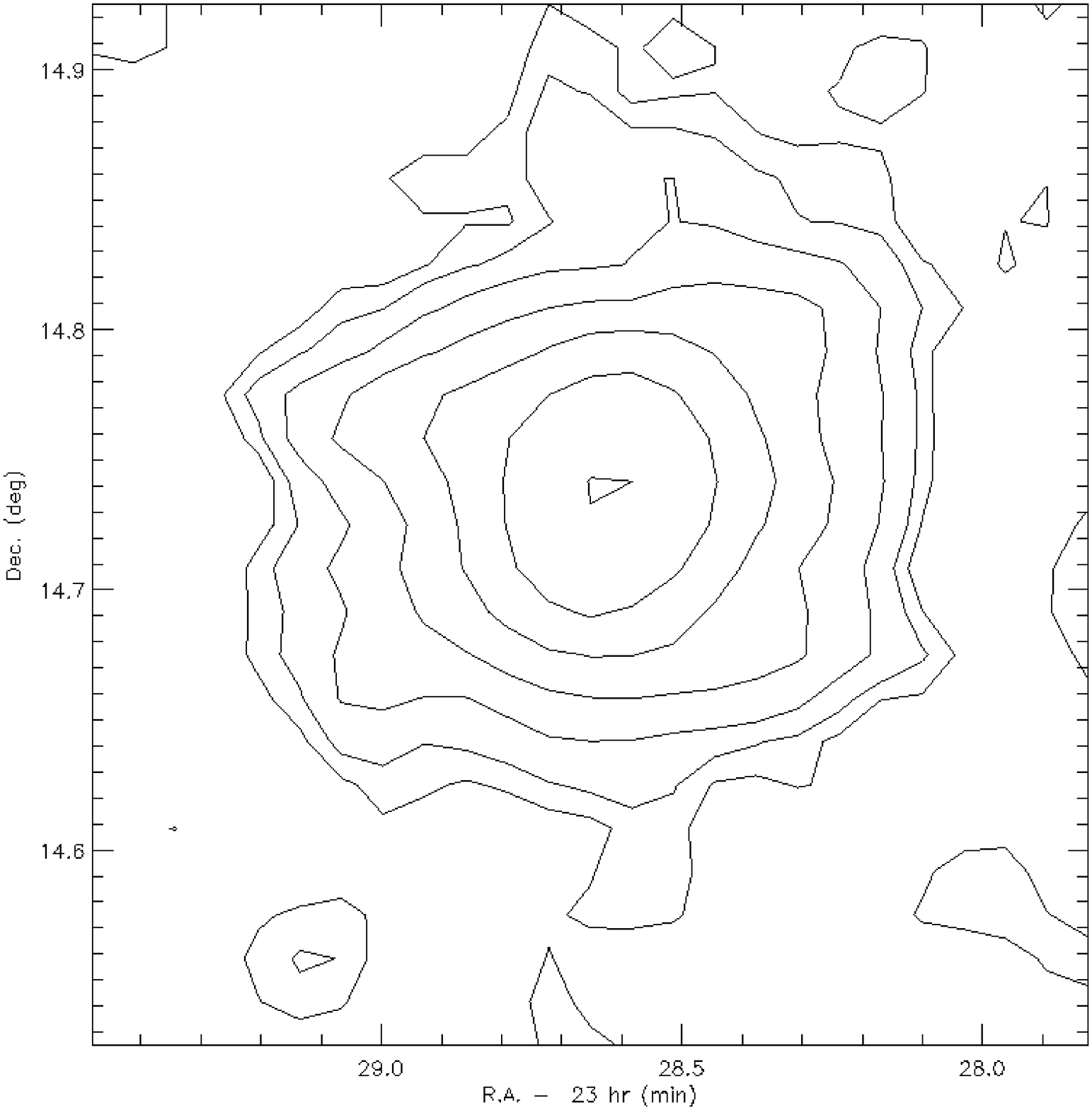}
    \includegraphics[angle=0,width=4.5cm]{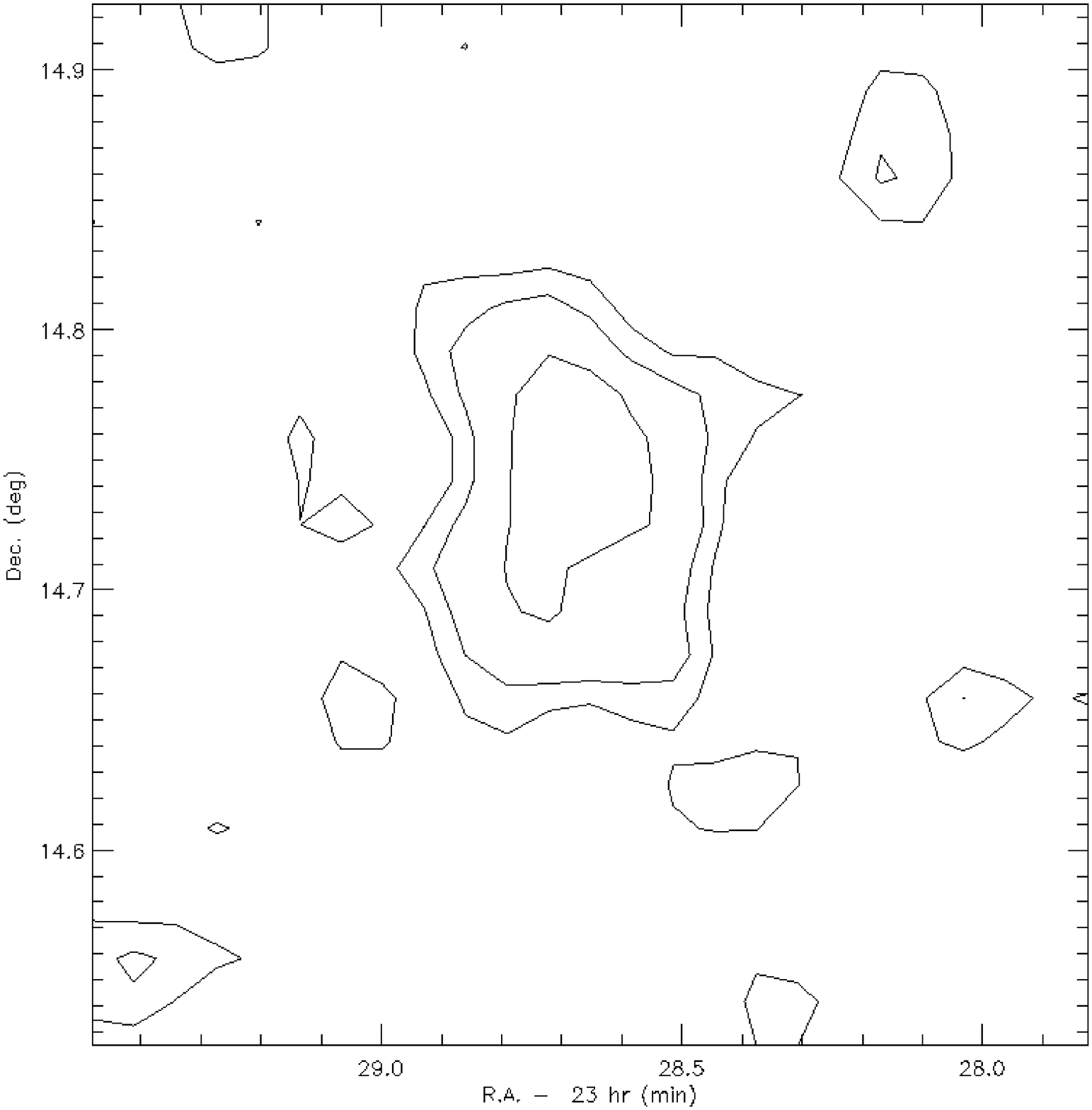}
    \includegraphics[angle=0,width=4.5cm]{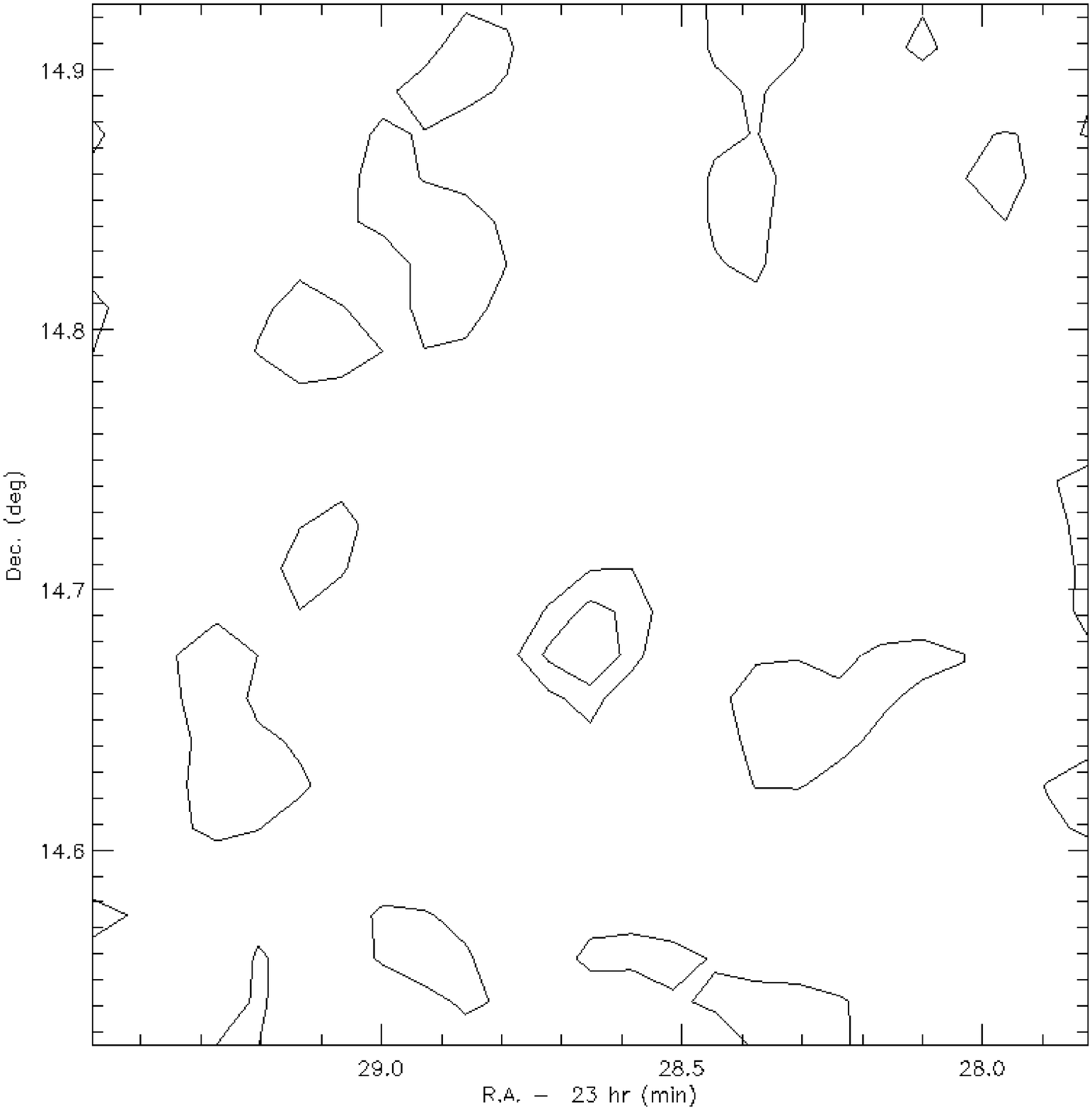}
    \includegraphics[angle=0,width=4.5cm]{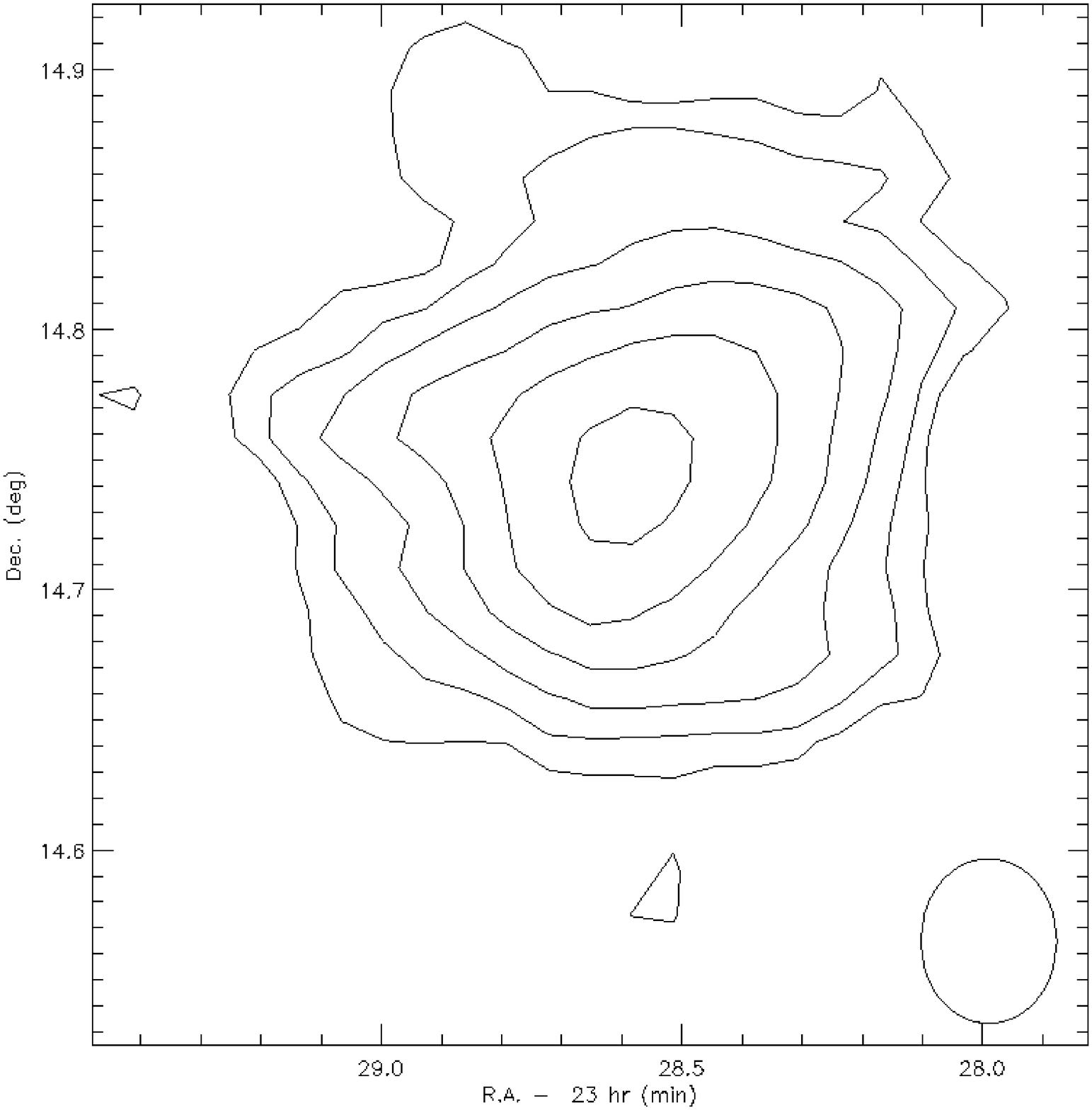}
        \caption{
Channel maps of the HI distribution and the integrated HI of PegDIG,
derived from the ALFALFA observations. The central velocities of each
channel map are (from left to right) $-138$, $-153$, $-169$ km s$^{-1}$ in the top row,
$-184$, $-200$, $-215$ km s$^{-1}$ in the middle row, and $-231$ km s$^{-1}$ in
the bottom row.
The total HI map is shown in the bottom-right panel, which also displays the
beam used to produce all the maps.
    \label{fig:HI_channel_maps}}
    \end{center}
\end{figure*}

The correction factors for the $gri$ filters were estimated visually,
using both the SBPs and the colour diagrams. Their final accuracy
is presumably about $\pm$0.1 mag, since the colour diagrams are
very sensitive to small photometric errors, and we estimate these
to be $\sim$0.05 mag for each band used here. The correction
factors themselves are 2.25 mag for $g$, 1.85 mag for $r$ and 1.65
mag for $i$ filter. They were obtained by shifting vertically the
values for the outer part of the galaxy to match those
for the inner, unresolved part of PegDIG.

\begin{figure}
    \begin{center}
    \includegraphics[angle=0,width=7.5cm]{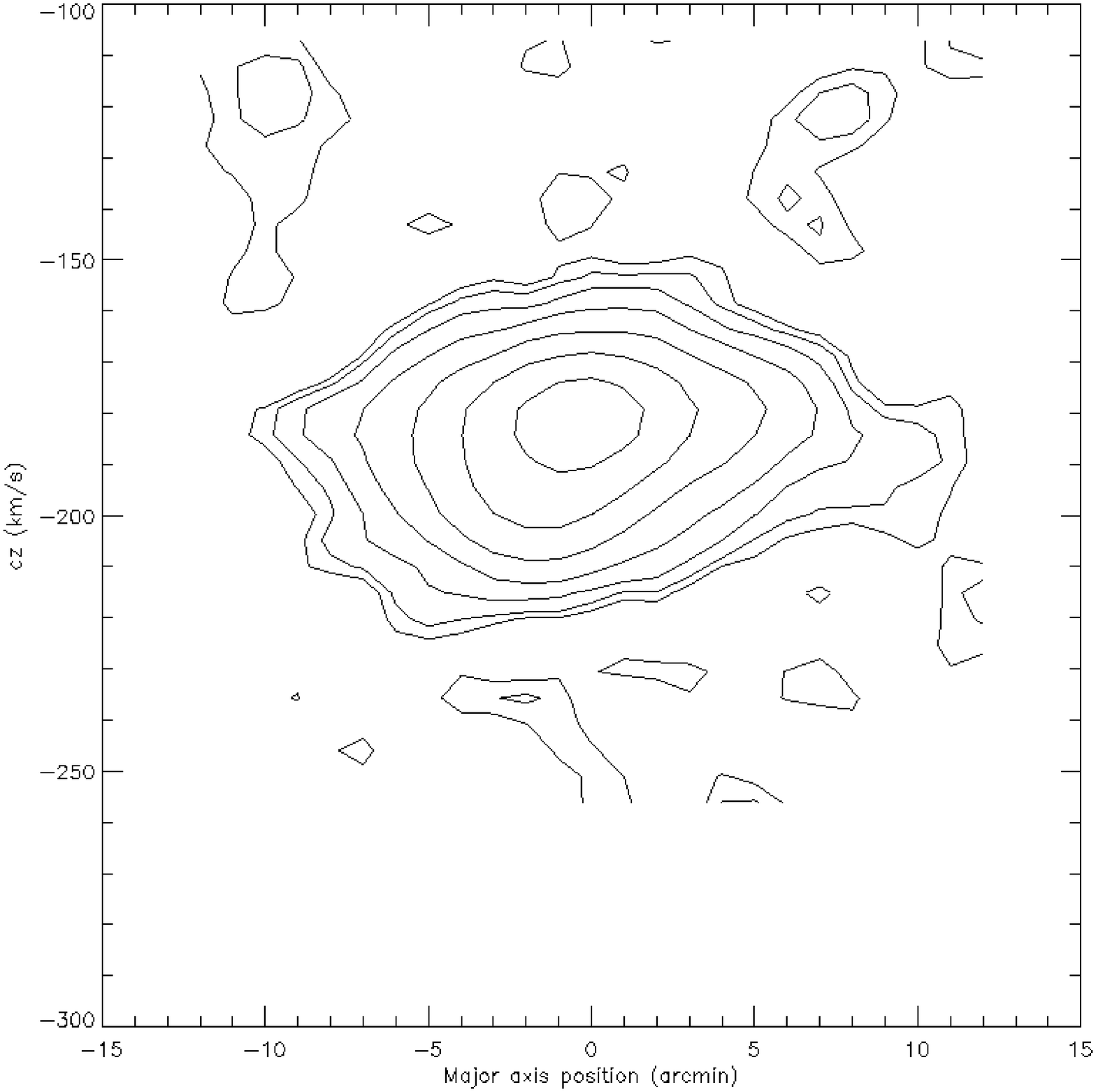}
    \includegraphics[angle=0,width=7.5cm]{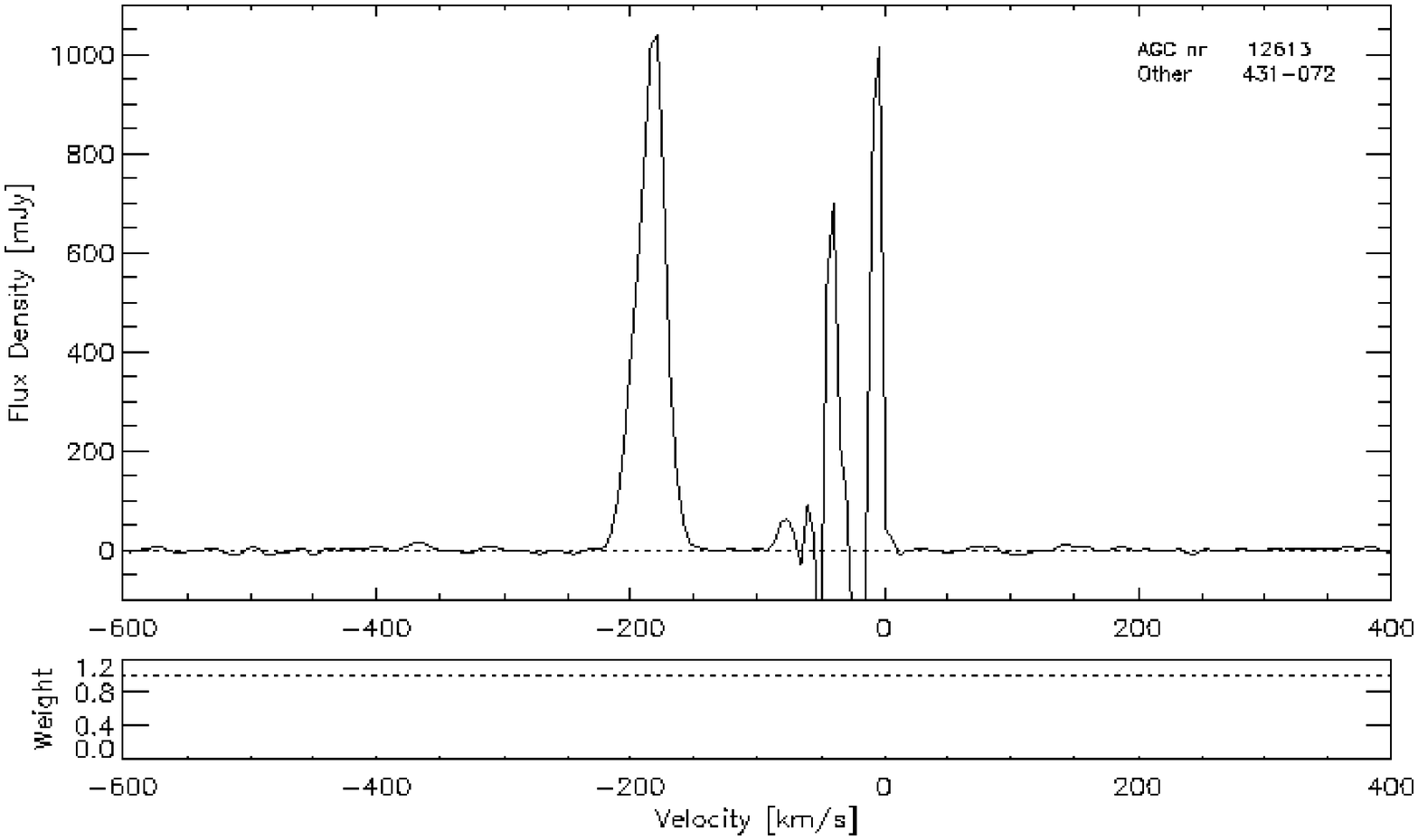}
        \caption{
	{\it Top panel}: Position-velocity contour map along the major axis of PegDIG.
	{\it Bottom panel}: The global spectrum (to the 100 mJy km s$^{-1}$ iso-intensity
	level) of PegDIG. The peaks between 0 and -100 km sec$^{-1}$ are Galactic hydrogen.
    \label{fig:HI_kinematics}}
    \end{center}
\end{figure}

\section{HI observations}
\label{txt:HI_obs}

ALFALFA, the Arecibo Legacy Fast ALFA extragalactic H\,{\sc i}
survey \citep{Giov05} is a blind neutral hydrogen survey that will ultimately
cover 7074 square degrees of the high galactic latitude sky visible from Arecibo.
It provides an extragalactic H\,{\sc i} line spectral database covering
the redshift range between $-1600$ km s$^{-1}$ and 18,000 km s$^{-1}$.
The ``fall'' part of the survey maps the sky region from RA=22$^h$ to 3$^h$
over most of the declination range accessible from Arecibo ($\delta$=0--36$^{\circ}$).
We collected the relevant HI observations covering PegDIG from the ALFALFA
archives as a grid containing the galaxy. The ALFALFA observations are conducted
at the Arecibo telescope with the 7-feed ALFA receiver. As described elsewhere
(Giovanelli et al. 2005a, 2005b, 2007), data acquisition for ALFALFA
is done in a fixed azimuth, drift-scan mode. Each surveyed sky region
is covered by two partly-overlapping drift scans collected at two different
epochs in order to sample at better than Nyquist spatial frequency.

The two scan passes are separated in time by several months, thus comparing the data
from the two epochs helps in ruling out spurious detections. When data acquisition
is completed over a given region of the sky, the individual drift scans are assembled
and regridded to form three-dimensional data cubes or ``grids''.
These grids are 2.4$^{\circ}\times$2.4$^{\circ}$ in size with 1 arcmin sampling,
and have 1024 channels along the spectral axis. Since the velocity resolution
is $\sim$5.3 km s$^{-1}$ (before Hanning smoothing), in order to cover
the full ALFALFA velocity range which extends from --2000 to 18,000 km s$^{-1}$
each grid is separated into four redshift subgrids, covering the velocity ranges:
$-2000$ -- 3300 km s$^{-1}$, 2500 -- 7900 km s$^{-1}$, 7200 -- 12,800 km s$^{-1}$,
and 12,100 -- 17,900 km s$^{-1}$. For mapping the HI in
PegDIG we used only the first velocity grid of the data cube at the PegDIG coordinates.

The central pixel of ALFA has low sidelobes, but the outer six pixels have
significant coma lobes. Furthermore, the contributions of
stray radiation and outer sidelobes to all the beams is not negligible.
Data at each gridpoint in the ALFALFA grids are obtained by a weighted average
of data from each beam that passes within a few arcmin from that point, thus the
effective beam at any point has a quite complex shape.
An IDL procedure for constructing effective beams for individual gridpoints
from the beammaps of the seven separate beams \citep{IHS+} was written by J. Dowell
\citep{GGC+}.
Using this procedure, we have constructed effective beams at each declination
pixel in the grid.
Two of these are shown in Figure~\ref{f:beams}.

The pronounced extensions shift in direction from one declination to another
as little as $2\arcmin$ away as the dominant contribution to the received
flux shifts from one ALFA pixel to the next.
At a level of 10\% of the main beam, these extensions to the effective beams
cause strong, extended sources like PegDIG to have apparent emission reaching
a few arcmin beyond the edge of the actual \ion{H}{I} distribution at any declination
where one of these extensions stretches toward the position of peak emission from
the galaxy.

Figure~\ref{fig:HI_channel_maps} shows the individual HI channel maps (averages
of 3 ALFALFA velocity channels each) at the velocities $-137.9$, $-153.4$, $-168.8$,
$-184.3$, $-199.7$, $-215.1$, and $-230.6$ km s$^{-1}$.  The contours are drawn at 2.15,
4.64, 10.0, 21.5, 46.4, 100, 215, 464, and 1000 mJy.
The bottom-right panel in Figure~\ref{fig:HI_channel_maps} presents
the total HI map (integrated over velocity) with contour levels
of 0.35, 0.74, 1.60, 3.5, 7.4 and 16.0 $\times10^{19}$ atoms cm$^{-2}$.
The ALFA beam used to produce these maps is plotted at the lower-right corner
of the total HI map; this is a fairly elliptical beam elongated north-south,
with major axes of 3'.3$\times$3'.9. The higher level contours agree well
with the previous map of \citet{HSF96} derived from Arecibo observations
with the flat feed. The lower contours in our map show some differences,
partly because the 1996 map was not a complete one with points spaced along
major and minor axes as opposed to a full grid, and because the side-lobe structure
is very different between the flat feed and the seven-feed ALFA.

From an inspection of the gridview display of the individual channels in
Figure~\ref{fig:HI_channel_maps}, we estimate that the structure in the
lowest two contours to the north of the galaxy is possibly, as explained above, emission into
the sidelobes of the beams that spanned that region and is probably spurious.
The extensions to the east and southwest could also be sidelobe emission
or could be real; this requires further observations.

The global spectrum (to the 100 mJy km s$^{-1}$ isophote level) for PegDIG,
derived from the ALFALFA observations, is shown in the left panel of
Figure~\ref{fig:HI_kinematics}.  The systemic HI velocity is $-183.6$ km s$^{-1}$
and the width at 50\% of the peak is 23.4$\pm$2.4 km s$^{-1}$.
The corresponding quantities at 20\% of the peak are $-185.2$ and
38.6$\pm$2.4 km s$^{-1}$.  The HI profile is best-fitted by a Gaussian centred
at $-183.4$ km s$^{-1}$ with FWHM=23.6$\pm$0.3 km s$^{-1}$.
The total HI flux is 28.1$\pm$0.1 Jy km s$^{-1}$, or 27.83$\pm$0.06 Jy km s$^{-1}$
if the integral under the best-fit Gaussian is used.
This corresponds to a total HI mass of $6.6\cdot 10^6$ M$_{\odot}$ at the 1 Mpc
nominal distance. There are no visible high-velocity HI wings.
Note also that the signal from PegDIG is clearly distinct from the
Galactic emission between $-100$ and 0 km s$^{-1}$ at this velocity resolution.

Figure~\ref{fig:HI_kinematics} shows the major axis position-velocity
(PV) contour map of PegDIG.  To produce it we rotated
the declination (DEC) -- right ascension (RA) maps by 25$^{\circ}$ (which we estimate
to be the angle of the major axis of the HI distribution away from the RA axis),
then summed along the minor axis.  The result is  plotted as a contour map
in the position vs.\ velocity plane, with contour
levels 21.544, 46.416, 100.0, 215.44, 464.16, 1000.0, and 2154.4 mJy.
The lowest two levels show the effects of the coma lobes, but the rest of
the contours should be robust. The PV map indicates a solid-body rotation curve
with opposite ends differing by 15 km s$^{-1}$, with a stream to more negative
{\it cz} about 5 arcmin from the centre of the galaxy.
The positive direction of the position axis is toward the SE.

We measured a flux integral similar to that of Young et al. (2003) and almost
twice that measured by Sill \& Israel (2002). We conclude that, to a level
of 2 mJy/beam, the galaxy is 0.3 degrees wide in RA and 0.25 degrees in DEC.
The peak of the HI is at RA = $23^h28^m55^s$, DEC $+14.75\degr$ in J2000 coordinates
and at about $-184$ km s$^{-1}$.
The FWHM of the HI line is about 28 km s$^{-1}$, about 10\% wider that the width
derived by Young et al.
The HI distribution does not show an obvious compression ridge to the SE,
or a tail to the NW smoothed out by our wide beam.
The slight crowding of the contours to the SE of the total
HI distribution and slight elongation to the NW may be
indicative of the compression ridge and tail inferred by
Young et al.\ (2003) from their VLA map, but lacking a
procedure to deconvolve the multiple asymmetric beams from
the map we cannot be certain.  However, we do not see any
extension of the HI significantly beyond the outermost
Young et al. contour even though ALFALFA is sensitive to
column densities 4 or 5 times lower than the sensitivity
limit of the Young et al. map.

\begin{figure*}
    \begin{center}
    \includegraphics[angle=-90,width=7.0cm]{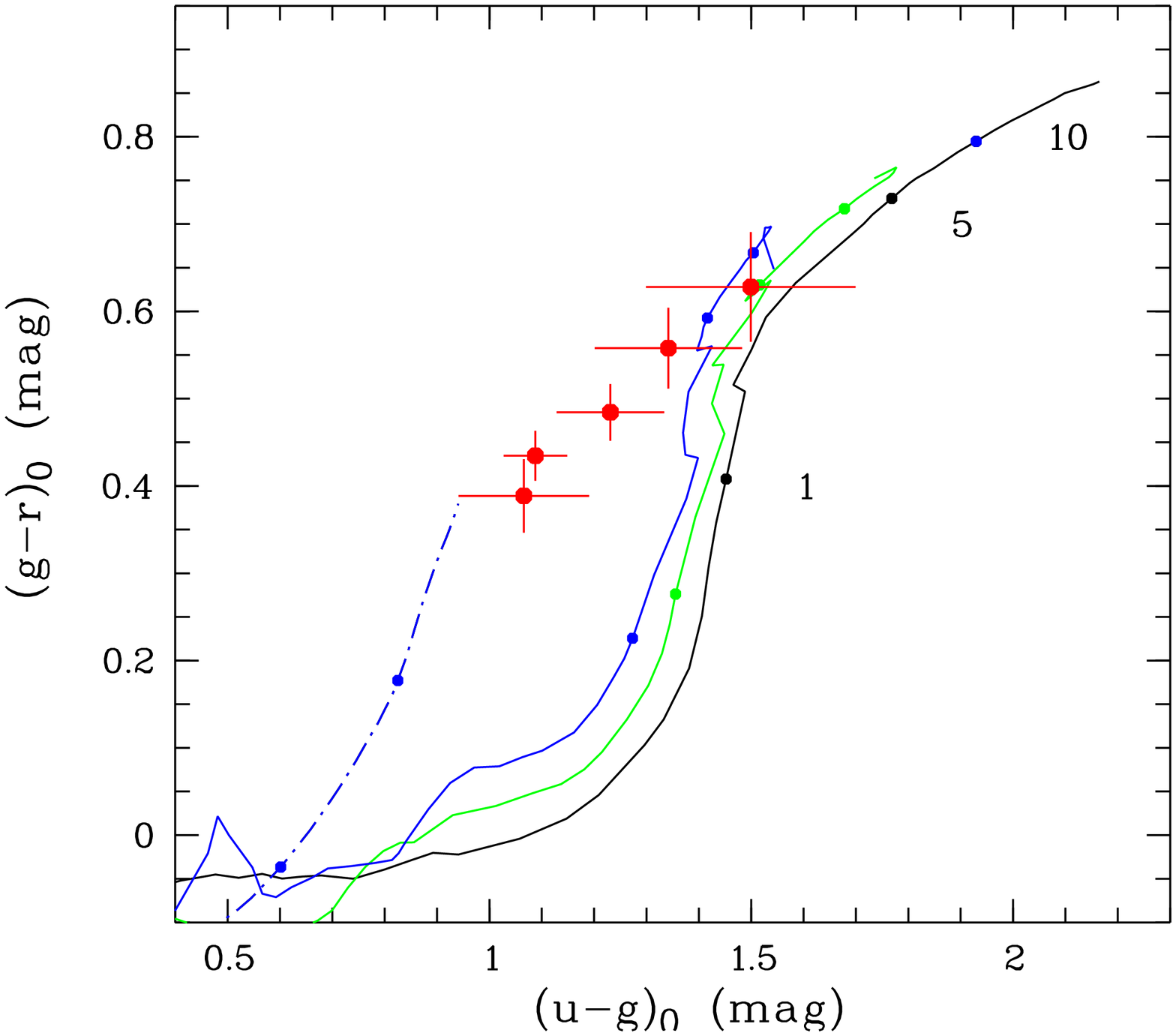}
    \includegraphics[angle=-90,width=7.0cm]{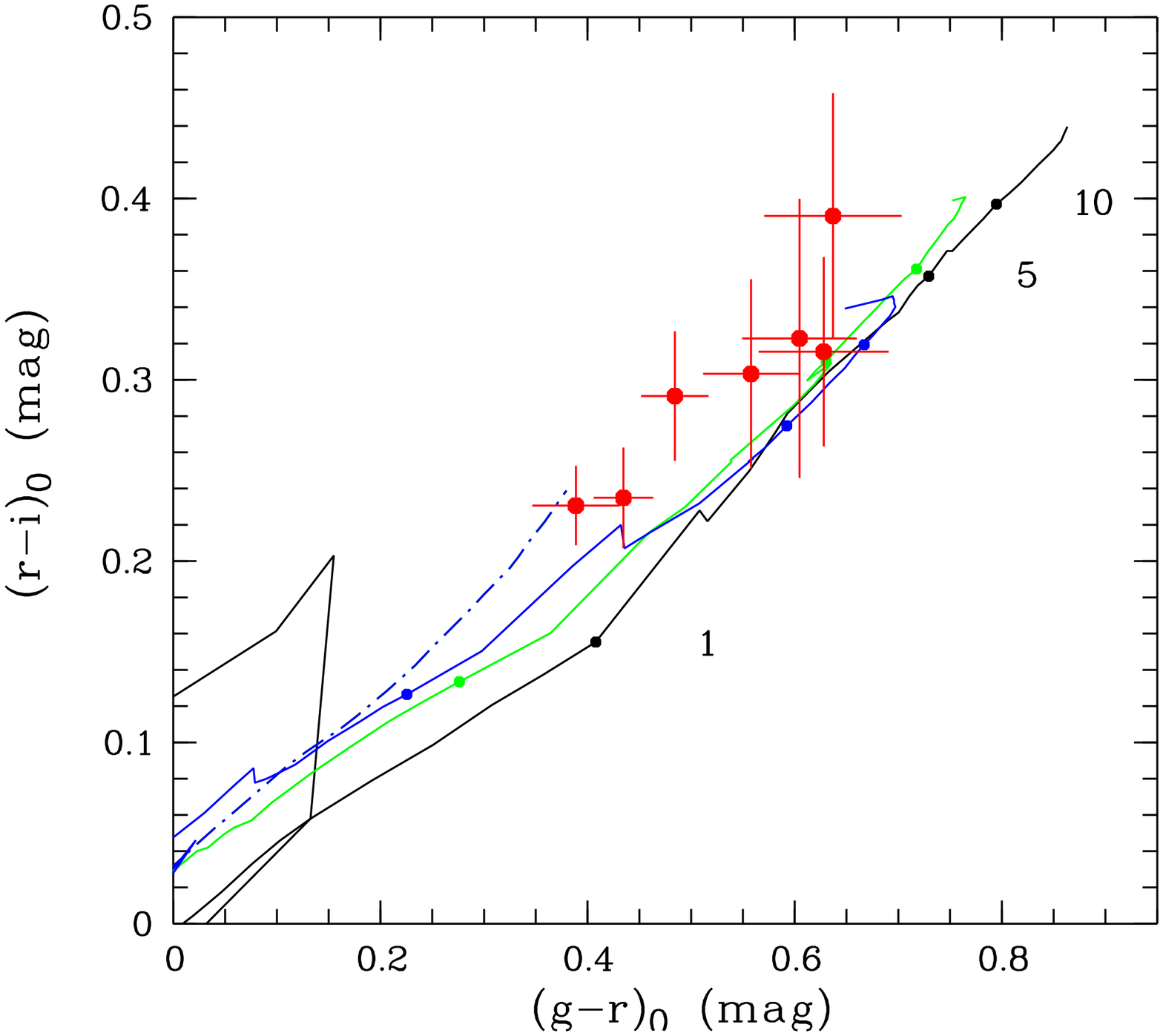}
    \includegraphics[angle=-90,width=7.0cm]{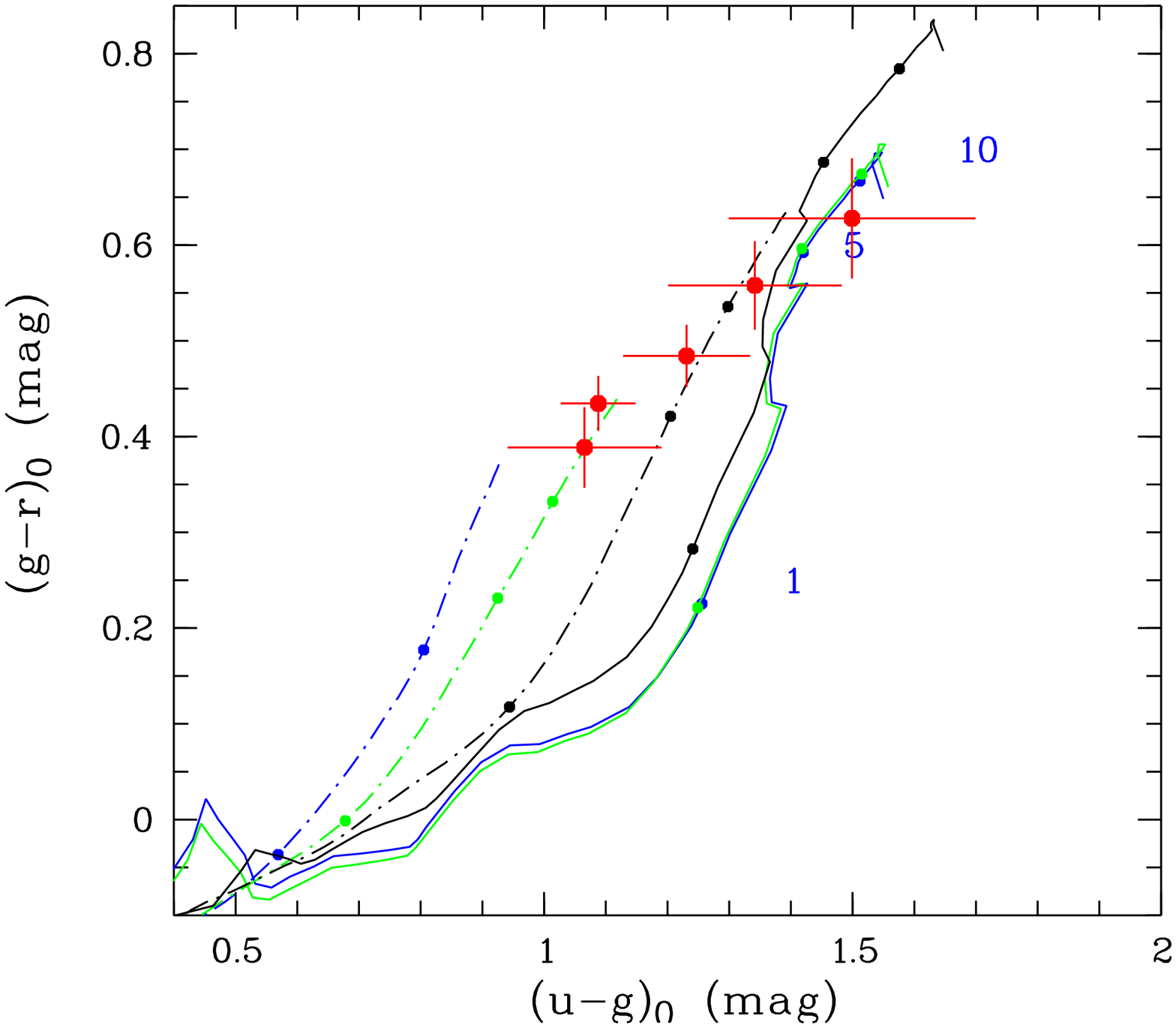}
    \includegraphics[angle=-90,width=7.0cm]{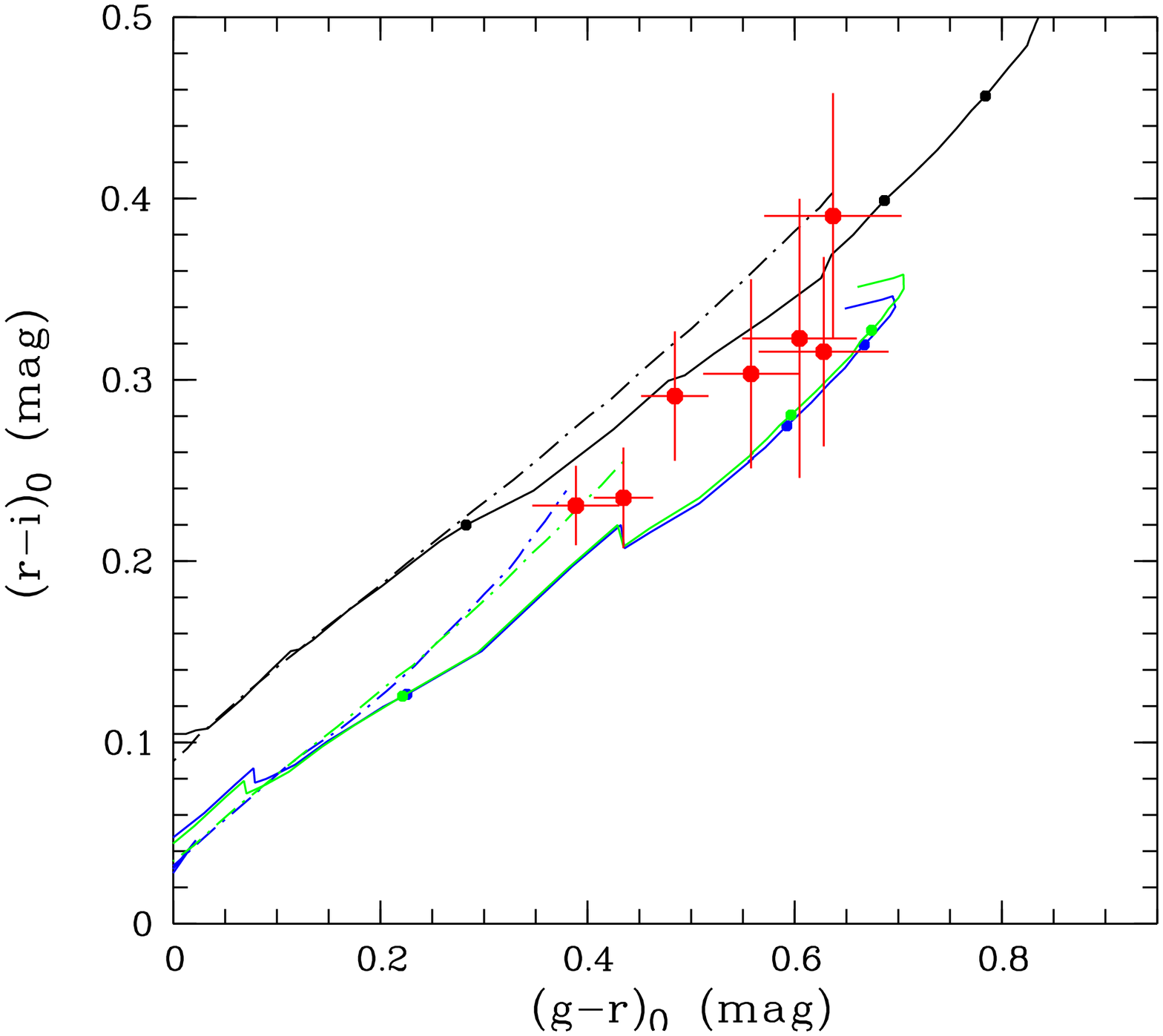}
    \caption{Two-colour diagrams with theoretical tracks from PEGASE2 for evolving
stellar populations. The observed colours are corrected for Milky Way foreground extinction
and are shown as red filled circles with error bars for the following major axis distances:
0--50, 50--100, 100--150, 150--250, 250--400, 400--600 and 600--800 arcsec.
$(u-g)_0$ could be only calculated up to a major axis distance of 400 arcsec.
The observed colours were averaged inside each major axis distance bin and the
$\pm 1 \sigma$ errors also reflect the scatter of colour in each distance range.
Redder colours correspond to an increased major axis distance: the lowest-leftmost
point represents the centre of PegDIG and highest-rightmost one corresponds to
the outermost part.
{\it Top panels:} Tracks with a standard Salpeter IMF, for instantaneous (solid)
and continuous (dashed line) SF laws using different metallicities: blue -- $z$=0.004
(1/5 Z$_{\odot}$), green -- $z$=0.008 (1/3 Z$_{\odot}$) and black -- Z$_{\odot}$.
Filled black hexagons along the tracks, with respective numbers, mark the time since
the beginning of SF: 1, 5 and 10 Gyr.
{\it Bottom panels:} Tracks for instantaneous (solid) and continuous (dashed line) SF laws
with a standard Salpeter IMF (blue) for $z$=0.004 and non-standard Salpeter IMFs:
slope is $-$1.6 for green and $-$2.85 for black.
    \label{fig:Model_colors}}
    \end{center}
\end{figure*}

\section{The stellar populations}

\citet{Zee00} presented $B$ and $(U-B)$, $(B-V)$ profiles that extend
only $\sim$300 arcsec from the centre; these show that the stellar
population changes from $(B-V)\simeq$0.6 one arcmin from the
centre ($\mu_B\simeq$24) to $\sim$0.8 some four arcmin from the
centre. This is similar to the colour gradient found here for
the unresolved part of the galaxy. 

Hierarchical galaxy formation scenarios predict that dwarf galaxies should
show traces of an old stellar population. And indeed, all nearby dwarf galaxies
show clear evidence for the presence of old stellar populations \citep{GreGal04},
i.e., populations older than 10 Gyr, as traced by, e.g., the presence of a horizontal
branch. This may also be the case for PegDIG, where traces of present-day very
low level star formation (4.4$\times$10$^{-5}$ M$_{\odot}$ yr$^{-1}$) were found
\citep[e.g.,][]{HE04} while most of the stars belong to earlier generations.
Indeed, the deepest existing HST data show clear evidence for substantial
populations with ages of 9 Gyr and younger as traced by a very prominent
red clump \citep{Gall98}. Those data approach the detection limit just below
the intermediate-age red clump, making it very difficult to establish
the presence of a horizontal branch
\citep[although there are indications of a density enhancement in the corresponding
region in the colour-magnitude diagram in Fig.\ 15 of][]{Gall98}.

\citet{Zee00} found that PegDIG had the reddest $(U-B)$ and $(B-V)$
colours of all the galaxies in her sample. She could not fit these
with a constant star formation model, but could fit them
with a continuous but declining star
formation rate, or with an aging, a few Gyr old starburst. She
proposed that the very low star formation rate, now relegated to
the innermost parts of the galaxy, might result from the
depletion of the galaxy's HI reservoir. \citet{Zee00} mentioned
also that the HII regions detected in her H$\alpha$ images are so
faint that they would not have been detected if PegDIG were more distant.

To estimate the distribution of stellar population ages in PegDIG, we compared
the derived integrated colours in its different parts with model tracks from
the PEGASE2 package \citep{pegase2} for metallicity values of
$z$=0.02 (solar), $z$=0.008 (1/3 of solar) and $z$=0.004 (1/5 of solar).
Since the photometric systems
($u^{\prime}$,$g^{\prime}$,$r^{\prime}$,$i^{\prime}$,$z^{\prime}$)
used for the calculations of PEGASE2 evolutionary tracks and
($u,g,r,i,z$) used in the real SDSS observations are slightly
different, we applied the transformation formulae from \citet{Tucker06}
in order to correct theoretical values to the ($u,g,r,i,z$) system.
PegDIG has been found to have a low present-day ISM metallicity of $\sim$1/5
of solar \citep{SBK97} and possibly has more metal-poor older stars \citep{Gall98}.
Based on the models we used, tracks of even lower metallicities ($<$1/5 of solar)
would not match the red colours
observed in the outer parts of PegDIG and thus these tracks are not plotted.

In Fig.~\ref{fig:Model_colors} we plot model tracks of colour
evolution in the $(g-r)_0$ versus $(u-g)_0$ and
$(r-i)_0$ versus $(g-r)_0$ diagram for a standard
Salpeter IMF, with lower and upper mass cutoff limits of 0.1 and 120
M\sunn. All observed colours are corrected for Milky Way foreground extinction.
Different tracks in Fig.\ \ref{fig:Model_colors}
represent the colour evolution for continuous SF with constant SFR
(dashed lines) and for an aging instantaneous SF episode (solid lines)
as two extremes of all possible SF histories. The hexagons on the
evolutionary tracks with the respective numbers mark ages in
Gyr since the beginning of SF.

We seek to constrain the mean age of the dominant population in different
annuli by trying to fit simultaneously both colour-colour plots, the
$(u-g)_0$ vs.\ $(g-r)_0$ and the $(g-r)_0$ vs.\ $(r-i)_0$ colours.  Clearly,
younger populations are more prominent in the central regions, where the mean
colours agree relatively well with the 1 Gyr tracks.  In the outermost regions
the averaged age is 5--10 Gyr or less.
Our data do not allow us to prove or to disprove the presence of an
old population (older than 10 Gyr), but the comparison of the integrated
colours with population synthesis models suggests that the contribution
of such an old population to the integrated light is fairly minor.
Hence the colour gradient appears to be
consistent with an age gradient in the sense that younger populations are
more centrally concentrated, while older populations show a more extended
distribution.  Of course, independent of the integrated colours, we know already
from published earlier studies that an age gradient is present,
as indicated by, e.g., the centralised distribution of the HII regions.

As is well-known, colour gradients may be caused by either age, metallicity,
or reddening gradients \citep{Harb01,Hunter06}. We cannot unambiguously disentangle
these three effects in our data, but we can at least attempt to qualitatively
comment on their importance. Since the intrinsic reddening of PegDIG seems to be
low \citep{Krienke01}, we discard extinction as a significant contributor to the
observed colour gradients. We have insufficient data to constrain a potential
metallicity gradient, but we note that the slope of the metallicity gradient
has to have an opposite sign to the observable one: the more metal-poor track has
bluer colours compared to the more metal-rich track.
In other dwarf galaxies for which more detailed data are
available, metallicity gradients tend to be small despite considerable
scatter at a given age \citep[e.g,][]{Sextans,Glatt08,Koch06}.
Altogether, it seems likely
that our large-scale colour gradients are indeed primarily driven by age,
although an age gradient may certainly also be linked with a metallicity gradient.

Population gradients have been identified in early-type and late-type dwarf galaxies alike
\citep[e.g,][]{Harb01,Parodi02,Makarova02,Makarova05,Lisker06,Hunter06},
although not all dwarf galaxies show such gradients. Generally seen
as colour gradients, such variations are usually interpreted as age gradients.
They appear to indicate longer-lasting star formation in the deeper, inner parts
of the potential well where the star-forming material accumulates more easily
and/or can be retained more easily.  PegDIG seems to fit this trend as well.

\section{Discussion}
\label{txt:Discussion}

One of the results of the present work is the derivation of surface brightness
profiles for PegDIG that extend about 1.6 times further out than what was found
in previous surface photometry studies \citep{Ti06}.
Other deep optical surveys of dIrrs in the Local Group have found additional evidence
for stellar distributions much more extended than previously
thought:  e.g., NGC\,6822 \citep{deBlok06}, Leo\,A \citep{Vans04},
and IC\,1613 \citep{Batt07}. \citet{Ti05,Ti06b} also found such an effect
for many nearby face-on and edge-on dIrrs.

Our deep surface photometry, using stars identified as belonging
to PegDIG and eliminating Milky Way stars in the foreground, showed that
it is possible to fit a S\'{e}rsic surface brightness distribution to the galaxy
at least to a radius of 14 arcmin$\simeq 4$ kpc. The stellar distribution
shows some concentrations around the unresolved 2 kpc part, and
some irregular extensions at the NW end of the major axis. Note
that this region shows less HI than the SE part, when our maps are
compared with those for neutral hydrogen from \citet{YvZL03}.
The HI column density shown in their Fig. 5 has a peak of
$\sim10^{21}$ atoms cm$^{-2}$ at RA = $23^h28^m35^s$, DEC= $+14^{\degr}44^m15^s$,
identical to the peak of the ALFALFA maps.

Our results allow the derivation of some general properties of PegDIG.
The total HI flux and the total B magnitude yield M(HI)/L$_B\simeq$0.4.
A rough estimate of the total dynamical mass can be derived from the PV plot
in the left panel of Figure~\ref{fig:HI_kinematics}.
Assuming that the outermost gaseous regions of PegDIG rotate with $\sim$60 km s$^{-1}$
at a galactocentric distance of 4 kpc, the total dynamical mass is
M$_{dyn}\simeq3.3\times10^8$ M$_{\odot}$. The neutral hydrogen contributes only
2\% of this mass. The mass in stars can be roughly estimated from the SDSS colours
plotted in Figure~\ref{fig:Model_colors}; these correspond to main-sequence K stars.
If the light were produced solely by K5 stars, the total mass in stars would
be M$_*=1\times10^8$ M$_{\odot}$, about 30\% of the total dynamical mass.
This indicates that while PegDIG has a significant amount of dark matter,
it is not a dark-matter-dominated galaxy, contrary to the assertion of \citet{AGB97}.

\citet{WKE04} interpreted the sharp drop in the surface brightness profile of the
Draco dSph as the signature of a kinematically cold stellar population in the outer
parts of the galaxy. They proposed two possible explanations for this phenomenon,
which they presented along with arguments against accepting them.
One is a two-population model with a hot bulge and a cold disk or halo that does
not fit either the observed kinematics nor the light distribution. The other is
tidal reshaping by the Milky Way galaxy, which requires very tight constraints on
the orbit of the dwarf galaxy. We mention this here since a similar phenomenon
may be present in PegDIG, even though PegDIG is more than five times as distant
from the nearest massive spiral as the Draco dSph \citep{GGH03}.

\citet{KKH04} mentioned that the major disturber galaxy for PegDIG, in terms of
tidal interactions, is M31, with a $\theta$ parameter of 1.2. This parameter is
the tidal index, which is defined as:
\begin{equation}
\theta_i=max[log \frac{M_k}{(D_{ik})^{3}}]+C
\end{equation}
Here $i$ is the index of the specific galaxy $\theta$ is
calculated for, $k$ is the index of any other galaxy, M$_k$ is the
mass of the $k$-th galaxy and D$_{ik}$ is the 3D space distance
between the $i$-th and the $k$-th galaxy \citep{KM99}. Galaxies with $\theta
\leq$0 can be considered to be undisturbed, while objects with
$\theta \geq$5 are considered to be highly disturbed. Based on the
calculated value, PegDIG could be somewhat affected by M31,
but not by an extreme tidal interaction. \citet{HE04}
listed M31 as the nearest neighbour to PegDIG, at a projected
distance of 450 kpc and a relative recession velocity of +117 km
sec$^{-1}$.  Note though that since the actual orbit of PegDIG is not known,
and it could even be on a fairly radial orbit around M31, a relatively strong
tidal interaction in the past cannot be excluded.

\citet{MVI07} plotted some of the PegDIG neighbours in
their Fig. 1 (lower right panel). Their argument is that a
comparison of the HI contours with the stellar distribution fits a
morphology of gas being stripped away by ram pressure. They
identify a ``compression front'' on the SE end of the
galaxy. In order for ram pressure stripping to take place, they
require the Local Group to be filled by a tenuous ($\sim10^{-5}$
cm$^{-3}$), rather warm ($\sim10^5$K) gaseous medium.
Ram pressure stripping would also modify the gas distribution
in and around the galaxy. In fact, this is the main argument used
by McConnachie et al.\ in bringing up the stripping possibility.
The morphological modifications that should appear if the ram pressure
stripping argument is valid are a leading edge compression region and
the creation of a tail of stripped material following the galaxy.

The deep HI maps shown in Figure~\ref{fig:HI_channel_maps} lack the signatures
of either a strong tidal interaction or of ram pressure stripping of the gas
from the galaxy. On the contrary, the low column density gas observed at the
outskirts of PegDIG and the lack of disturbances in the distribution of this
tenuous gas imply that any presumed interaction with an intergalactic medium,
as proposed by McConnachie et al (2007), can probably be discounted.
\citet{GGH03} presented simple estimates of the efficiency of ram pressure
stripping by a homogeneous Local Group intergalactic medium and concluded that the
densities are too low to have a noticeable effect.  However, \citet{GGH03}
also argued that a putative {\em clumpy} medium could be rather effective.  But the
absence of disturbances in the PegDIG HI contours does not support stripping
by a clumpy intergalactic medium.

With the present star formation rate derived by \citet{YvZL03}
and the total HI mass measured in this work, the star formation could last
for another 2$\times10^{11}$ years.  Such slow, continuous star
formation extending over a Hubble time or more is typical for dIrr galaxies
\citep[e.g,][]{Hunter86,Zee01}.  The central HI column density is a bit short
of the canonical threshold for star formation, but this canonical limit is not
always valid for low-mass galaxies \citep[e.g.,][]{Hunter98}.
Is it possible that we are seeing PegDIG during an extended lull in star formation,
while the gas is settling  back after having been distended by,
e.g., supernovae in the last star formation episode,
akin to scenarios described by, e.g., \citet{Dong03}?
Amplitude variations in the intensity of star formation are commonly
observed in dIrr galaxies, so here also PegDIG would conform to the typical
properties of these objects \citep[e.g.,][]{Tosi91,Greb97}.

\citet{BCK06} discussed the SF threshold in very faint low-mass
galaxies based on synthesis observations with the Giant Metrewave Radio Telescope
(GMRT). They found that while current star formation (as traced by H$\alpha$
emission) is confined to regions with relatively large HI column density
[N$_{HI}>$(0.4-1.7)$\times10^{21}$ cm$^{-2}$],
the morphology of the H$\alpha$ emission is generally not correlated with
that of the high HI column density gas. A high gas column density may be
a necessary condition for star formation, but it is not sufficient, for their
sample at least, to ensure that star formation does in fact occur.

We can also rule out tidal deformation, since such distortions are expected
to be symmetric with respect to the centre, while for PegDIG any distorsions
are relegated to the NW side.
Our findings imply that PegDIG might be a recent acquisition of the Local Group,
now in the outskirts of the LG and far away from any nearby massive galaxies,
that was formed in a comparatively empty region without major disturbers.
However, without knowledge of its orbit we cannot rule out past interactions with M31.

\section{Conclusions}
\label{Summary}

We analysed images of the Pegasus dwarf irregular galaxy collected
by the SDSS survey and found that the unresolved part can be
fitted by a S\'{e}rsic intensity profile down to $\sim30$ mag
arcsec$^{-2}$. Using very effective filtering in the colour-magnitude space
of SDSS data, we reduced the contamination by foreground Galactic
field stars and significantly increased the contrast for the outer
part of the Pegasus dwarf where we identified resolved stars that belong to PegDIG.
This allowed the extension of the surface photometry to much fainter levels.

Our extended surface photometry, reaching down to a surface
brightness of $\mu_g\simeq$33 mag arcsec$^{-2}$, revealed a $\sim$8
kpc wide stellar distribution following the same S\'{e}rsic profile as
found for the unresolved part, composed of a stellar
population similar to that in the $\sim$2 kpc main body, but significantly
older. The distribution of colours across the galaxy body shows that the
innermost parts of the galaxy contain the youngest population.
A comparison to population synthesis models suggests a mean age of
$\sim 1$ Gyr for this part, and we know from earlier work at HST
resolution that the youngest stars have ages of only a few 10 Myr.
The outermost parts of PegDIG are much older with a mean age from integrated colours
of at least 5 Gyr, indicating that there the contribution
of younger populations is comparatively small.
The total dynamical mass of the galaxy is $\sim3\times10^8$ M$_{\odot}$,
of which about 30\% is in stars and only 2\% is in HI.

We found that the stellar distribution of PegDIG is considerably more
extended than previously thought. Mapping the HI distribution to very low column
density levels at Arecibo revealed that the hydrogen distribution is slightly smaller
than that of the stars revealed by the present study. PegDIG is therefore yet another dIrr
where the HI is not much more extended than the stellar distribution,
as it used to be in the classical picture of dIrrs.

Our deep HI map shows an extended and fairly regular gas
distribution with solid-body-like rotation.  We do not observe
low column density tails extending beyond the edges of
the neutral gas distribution shown in earlier synthesis array
images (Young et al.\ 2003).  Tidal stripping therefore seems
unlikely.  On the basis of the HI observations alone, we cannot
rule out ram pressure interaction with extragalactic gas,
and the relatively small extent of the HI vis-a-vis the
distribution of stars may even support the hypothesis that
the outermost gas has been stripped.

We identified faint extensions of the optical light distribution of PegDIG
at the north-west end that do not follow the expected distortions caused by
a tidal interaction. We also showed that a number of stellar concentrations -- possibly
extended stellar associations -- are located around the unresolved central galaxy body.
Rings of stellar associations have been found in a number of dIrrs and could
be a possible sign of outward-propagating star formation but, on the other hand,
PegDIG also has very young stars in its innermost regions as shown by the HST
data \citep{Gall98}.

\section*{Acknowledgments}
\label{Thanks}

We thank the anonymous referee for comments that improved the
presentation of the manuscript.
This paper makes extensive use of SDSS data
products.  Funding for the SDSS and SDSS-II has been provided by the Alfred P.
Sloan Foundation, the Participating Institutions, the National Science
Foundation, the U.S. Department of Energy, the National Aeronautics and Space
Administration, the Japanese Monbukagakusho, the Max Planck Society, and the
Higher Education Funding Council for England. The SDSS Web Site is
http://www.sdss.org/.

The SDSS is managed by the Astrophysical Research Consortium for the
Participating Institutions. The Participating Institutions are the American
Museum of Natural History, Astrophysical Institute Potsdam, University of
Basel, University of Cambridge, Case Western Reserve University, University of
Chicago, Drexel University, Fermilab, the Institute for Advanced Study, the
Japan Participation Group, Johns Hopkins University, the Joint Institute for
Nuclear Astrophysics, the Kavli Institute for Particle Astrophysics and
Cosmology, the Korean Scientist Group, the Chinese Academy of Sciences
(LAMOST), Los Alamos National Laboratory, the Max-Planck-Institute for
Astronomy (MPIA), the Max-Planck-Institute for Astrophysics (MPA), New Mexico
State University, Ohio State University, University of Pittsburgh, University
of Portsmouth, Princeton University, the United States Naval Observatory, and
the University of Washington.

This work is based in part on observations collected at Arecibo Observatory.
The Arecibo Observatory is part of the National
Astronomy and Ionosphere Center, which is operated by
Cornell University under a cooperative agreement with the
National Science Foundation.

This research has made use of
the NASA/IPAC Extragalactic Database (NED) which is operated by the
Jet Propulsion Laboratory, California Institute of Technology, under
contract with the National Aeronautics and Space Administration.
Moreover, this research has made extensive use of NASA's Astrophysics
Data System.


\bsp

\label{lastpage}
\end{document}